\documentclass[natbib]{svjour3}
\journalname{Celestial Mechanics and Dynamical Astronomy}
\def\mearth{{\rm\,M_\oplus}}
\def\rearth{{\rm\,R_\oplus}}
\def\msun{{\rm\,M_\odot}}

\def\gsim{~\rlap{$>$}{\lower 1.0ex\hbox{$\sim$}}}
\def\lsim{~\rlap{$<$}{\lower 1.0ex\hbox{$\sim$}}}

\def\eg{{\it e.g.\ }}

\def\ie{{\it i.e.\ }}

\def\tess{{\it TESS}}
\def\kepler{{\it Kepler}}
\def\eqtide{\texttt{EQTIDE}}

\usepackage{graphicx}
\usepackage{mathtools}

\begin{document}

\title{Tidal Locking of Habitable Exoplanets}
\author{Rory Barnes}
\institute{Astronomy Department, University of Washington, Box 951580, Seattle, WA 98195 \and NASA Astrobiology Institute -- Virtual Planetary Laboratory Lead Team, USA}

\maketitle

\begin{abstract}
Potentially habitable planets can orbit close enough to their host
star that the differential gravity across their diameters can produce
an elongated shape. Frictional forces inside the planet prevent the
bulges from aligning perfectly with the host star and result in
torques that alter the planet's rotational angular
momentum. Eventually the tidal torques fix the rotation rate at a
specific frequency, a process called tidal locking. Tidally locked
planets on circular orbits will rotate synchronously, but those on
eccentric orbits will either librate or rotate
super-synchronously. Although these features of tidal theory are
well-known, a systematic survey of the rotational evolution of
potentially habitable exoplanets using classic equilibrium tide
theories has not been undertaken. I calculate how habitable planets
evolve under two commonly-used models and find, for example, that one
model predicts that the Earth's rotation rate would have synchronized
after 4.5 Gyr if its initial rotation period was 3 days, it had no
satellites, and it always maintained the modern Earth's tidal
properties. Lower mass stellar hosts will induce stronger tidal
effects on potentially habitable planets, and tidal locking is possible
for most planets in the habitable zones of GKM dwarf stars. For fast rotating planets,
both models predict eccentricity growth and that circularization can
only occur once the rotational frequency is similar to the orbital
frequency. The orbits of potentially habitable planets of very late M
dwarfs ($\lsim~0.1~\msun$) are very likely to be circularized within 1
Gyr and hence those planets will be synchronous rotators. Proxima b is
almost assuredly tidally locked, but its orbit may not have
circularized yet, so the planet could be rotating super-synchronously
today. The evolution of the isolated and potentially habitable {\it
  Kepler} planet candidates is computed and about half could be
tidally locked. Finally, projected {\it TESS} planets are simulated
over a wide range of assumptions, and the vast majority of all cases
are found to tidally lock within 1 Gyr. These results suggest that the
process of tidal locking is a major factor in the evolution of most of
the potentially habitable exoplanets to be discovered in the near
future.
\end{abstract}

\section{Introduction}
\label{sec:intro}

The role of planetary rotation on habitability has been considered for
well over a century. By the late 1800's astronomers were keenly
interested in the possibility that Venus could support life, but
(erroneous) observations of synchronous rotation led to considerable
discussion of its impact on planetary habitability
\citep{Schiaparelli1891,Lowell1897,Slipher1903,See1910,Webster1927}.
Some researchers asked ``[w]ill tidal friction at the last put a stop
to the sure and steady clockwork of rotation, and reduce one
hemisphere to a desert, jeopardizing or annihilating all existence [of
  life on Venus]?''  \citep{Mumford1909}, while other, more optimistic,
scientists suggested ``that between the two separate regions of
perpetual night and day there must lie a wide zone of subdued
rose-flushed twilight, where the climatic conditions may be well
suited to the existence of a race of intelligent beings''
\citep{Heward1903}. We now know that Venus is covered with thick clouds
and that astronomers misinterpreted their results, but their
speculations on the habitability of tidally locked worlds are similar
to modern discussions.

The possibility that the rotational period of some habitable
exoplanets may be modified by tidal interaction with their host stars
was first suggested by Stephen Dole in his classic book {\it Habitable
Planets for Man} over 50 years ago \citep{Dole64}. At the time, no
exoplanets were known, but Dole, motivated by the dawning of the
``space age," was interested in the possibility that humanity could
someday travel to distant star systems. \cite{Dole64} did not compute
planetary spin evolution, but instead calculated the heights of tidal
bulges in the Solar System to identify a critical height that
separates synchronously and freely rotating worlds. Citing the models
of \cite{Webster1925}, he settled on $\sqrt{2}$ feet (= 42~cm), and
then calculated the orbital distances from a range of stellar hosts
for which Earth's tide reached that height. Assuming (somewhat
arbitrarily) an ``ecosphere" of our Solar System to lie between 0.725
and 1.24 AU, he concluded that all potentially habitable planets
orbiting stars less than 72\% the mass of the Sun would rotate
synchronously and that the inner edge of the ecosphere could be
affected up to 88\%. As Dole believed it was ``evident that low rates
of rotation are incompatible with human habitability requirements," a
sense of pessimism developed regarding the possibility that planets
orbiting M dwarfs could support life.

\cite{Kasting93} returned to the problem and computed explicitly the
orbital radius at which a planet similar to Earth would become
a synchronous rotator around main sequence stars. In particular they
used a model in which the phase lag between the passage of the
perturber and the tidal bulge is constant and concluded that an
Earth-like planet's rotational frequency would synchronize with the
orbital frequency in the habitable zone (HZ; the shell around a star for which runaway
atmospheric feedbacks do not preclude surface water) for stellar masses $M_* < 0.42~\msun$ within
4.5 Gyr. They called the orbital distance at which this state
developed the ``tidal lock radius." They relied on the model of
\cite{Peale77} and used a relatively low energy dissipation rate for
Earth, as suggested by models of the evolution of the Earth-Moon
system in isolation \citep{MacDonald64}, see also
$\S$~\ref{sec:methods}.4. Furthermore, \cite{Kasting93} used a
sophisticated one-dimensional photochemical-climate model to compute
the HZ, which is more realistic than Dole's ecosphere.

The study of \cite{Kasting93} differs from \cite{Dole64} in that the
former was interested in planets which could support liquid surface
water, whereas the latter imagined where modern humans would feel
comfortable. It is important to bear in mind that synchronous rotation
represents a state for which the atmospheric modeling approach of \cite{Kasting93} breaks
down -- a planet with a permanent dayside and permanent nightside is
not well-represented by a one-dimensional model in altitude -- not a
fundamental limit to the stability of surface water on an Earth-like
planet. Nonetheless, \cite{Kasting93} retained Dole's pessimism and
concluded that ``all things considered, M stars rank well below G and
K stars in their potential for harboring habitable planets."  Clearly
the meaning of ``habitable" is important in this discussion, and
has affected how scientists interpret their results in terms of the search for
life-bearing worlds. In this study, a``habitable planet" is one that
is mostly rock with liquid water oceans (that may be global)
and a relatively thin ($\lsim~1000$~bars) atmosphere.

With the development of 3-dimensional global climate models (GCMs),
the surface properties of synchronously rotating planets can be
explored more self-consistently. The first models were relatively
simple, but found that synchronously rotating planets can support
liquid water and hence should be considered potentially habitable
\citep{Joshi97}. More recent investigations have confirmed this result
\citep{Wordsworth11,Pierrehumbert11,Yang13,Way15,Shields16,Kopparapu16},
and hence we should no longer view synchronous rotation as a limit to
planetary habitability. Moreover, these GCM models suggest the HZ for
synchronous rotators may extend significantly closer to the host star
than 1-D models predict \citep{Yang13}. These are precisely the planets
likely to be discovered by the upcoming Transiting Exoplanet Survey Satellite (\tess) mission
\citep{Ricker14,Sullivan15}, and may also be amenable to transit
transmission spectroscopy by the {\it James Webb Space
  Telescope}. Furthermore, low-mass stars are more common than Sun-like
stars and so a significant number of exoplanets in the HZ of nearby
stars may be in a synchronous state and with a rotational axis nearly
parallel with the orbital axis \citep{Heller11}.

Some recent studies have examined the rotational evolution of planets
and satellites \citep{FerrazMello15,Makarov15}, but did not consider
their results in relation to the HZ. Motivated by the potential to
detect extraterrestrial life, I have performed a systematic study of
the tidal evolution of habitable planets to provide more insight into
the physical and orbital properties that can lead to significant
rotational evolution and in some cases synchronous rotation. This
survey focuses solely on the two-body problem and neglects the role of
companions and spin-orbit resonances, both of which could
significantly affect the tidal evolution,
\eg~\citep{WuGoldreich02,MardlingLin02,Rodriguez12,vanLaerhoven14}. In
particular, I will use the ``equilibrium tide" (ET) model, first
proposed by \cite{Darwin1880} and described in more detail in
$\S$~\ref{sec:methods}, to simulate the orbital and rotational
evolution of rocky planets with masses between 0.1 and 10~$\mearth$
orbiting stars with masses between 0.07 and 1.5~$\msun$ for up to 15
Gyr.

\cite{Kasting93} employed an ET model and made many assumptions and
approximations, but it was vastly superior to the method of Dole. The
Kasting et al. model used a $1~\mearth,~1~\rearth$ planet with zero
eccentricity, no companions, no obliquity, and an initial rotation
period of 13.5 hours to compute the orbital radius at which tidal
effects cause the planet to become a synchronous rotator. In the ET
model of \cite{Peale77}, the rate of tidal evolution scales linearly
with the so-called ``tidal quality factor" $Q$, which essentially
links energy dissipation by friction with the torques due to
asymmetries in the bodies. \cite{Kasting93} chose $Q=100$ as suggested
by \cite{MacDonald64} despite the fact that modern measurements based
on lunar laser ranging, see \cite{Dickey94}, find that Earth's value
is $12~\pm~2$ \citep{Williams78}. \cite{Kasting93} also ignored 
the ET models' predictions that large eccentricities ($e~\gsim~0.1$) could
result in supersynchronous rotation
\citep{Goldreich66,Barnes08,FerrazMello08,Correia08}. ET models are not
well-calibrated at large eccentricities, but given the large number of
large exoplanets with large eccentricities, it may be that many
planets are ``tidally locked," meaning their rotation rate is fixed by
tidal torques, yet they do not rotate synchronously.
 
The method of calculating HZ boundaries of \cite{Kasting93} has been
improved several times, \eg~\citep{Selsis07,Kopparapu13}, and these
studies typically include a curve that is similar to that in
\cite{Kasting93} which indicates that rotational synchronization is
confined to the M spectral class. Such a sharp boundary in tidal
effects is misleading, as the initial conditions of the rotational and
orbital properties can span orders of magnitude, the tidal dissipation
rate is poorly constrained, and the tidal models themselves are poor
approximations to the physics of the deformations of planetary
surfaces, particularly those with oceans and continents. We should
expect a wide range of spin states for planets in the HZ of a given
stellar mass, an expectation that is borne out by the simulations
presented below in $\S$~\ref{sec:results}.

The ET model ignores many phenomena that may also affect a planet's
rotational evolution such as atmospheric tides
\citep{GoldSoter69,CorreiaLaskar01,CorreiaLaskar03,Leconte15}, the
influence of companions, \eg~\citep{Correia13,Greenberg13}, rotational
braking by stellar winds \citep{Matsumura10,Reiners14}, etc. These
effects can be significant, perhaps even dominant, and recently some
researchers have improved on ET models by including more realistic
assumptions about planetary and stellar interiors,
\eg~\citep{Henning09,Correia06,FerrazMello13,Zahnle15,DriscollBarnes15}. However,
tidal dissipation on Earth occurs primarily in the oceans as a result
of nonlinear processes and is not well-approximated by assumptions of
homogeneity. Satellite data reveal that
about two-thirds of Earth's dissipation occurs in straits and shallow
seas and about one-third in the open ocean \citep{EgbertRay00}. The
former represent bottlenecks in flow as the tidal bulge passes and
turbulence leads to energy dissipation. The latter is probably due to
seafloor topography in which gravity waves are generated by ocean
currents passing over undersea mountain ranges, leading to non-linear
wave interactions that cause energy dissipation. Thus, the tidal
braking of an Earth-like planet is most dependent on the the unknown
properties of a putative ocean. As no ``exo-oceanography" model has
been developed, the ET models seem reasonable choices to
explore the timescales for tidal braking of habitable exoplanets, but
with the caveat that they are not accurate enough to draw robust
conclusions in individual cases, rather they should only be used to
identify the range of possible behavior. Also note that an implication
of \cite{EgbertRay00}'s result is that exoplanets covered completely
by liquid water will likely only be a few times less dissipative than
one with a dichotomous surface like Earth.

This study focuses on the behavior predicted by the ET model over a
wide range of initial conditions using two popular
incarnations. Consideration of this broader range of parameter space
reveals that nearly any potentially habitable planet could have its
rotation period affected by tides. About half of \kepler's isolated
planet candidates are tidally locked if they possess Earth's tidal
properties, and Proxima b is found to have a tidal locking time of
less than $10^6$ years for all plausible assumptions. I also calculate
the time to tidally lock, $T_{lock}$ for the projected yields of
NASA's upcoming \tess~mission and find that in almost all cases
$T_{lock} < 1$~Gyr, suggesting all potentially habitable worlds it
detects will have had their rotations modified significantly by tidal
processes. I also examine how eccentricity can grow in some cases in
which a habitable planet is rotating super-synchronously, a process
that has been briefly discussed for non-habitable worlds
\citep{Heller10,Cheng14}. In most cases, eccentricity growth is modest,
and primarily acts to delay circularization. In
$\S$~\ref{sec:discussion} I discuss these results in terms of
exoplanet observations in general, and the possible impact on
planetary habitability.

\section{The Equilibrium Tide Model\label{sec:methods}}

The ET model assumes the gravitational force of the tide-raiser
produces an elongated (prolate) shape of the perturbed body and that
its long axis is slightly misaligned with respect to the line that
connects the two centers of mass. This misalignment is due to
dissipative processes within the deformed body and leads to a secular
evolution of the orbit and spin angular momenta. Furthermore, the bodies are assumed to respond to the
time-varying tidal potential as though they are damped, driven
harmonic oscillators, a well-studied system. As described below, this
approach leads to a set of 6 coupled, non-linear differential
equations, but note that the model is linear in the sense that there
is no coupling between the surface waves that sum to the equilibrium
shape. A substantial body of research is devoted to tidal theory,
\citep[\eg][]{Darwin1880,GoldreichSoter66,Hut81,FerrazMello08,Wisdom08,EfroimskyWilliams09,Leconte10},
and the reader is referred to these studies for a more complete
description of the derivations and nuances of ET theory. For this
investigation, I will use the models and nomenclature of
\citep{Heller11}, which are presented below.

ET models have the advantage of being semi-analytic, and hence can be
used to explore parameter space quickly. They effectively reduce the
physics of the tidal distortion to two parameters, which is valuable
in systems for which very little compositional and structural
information is known, \eg exoplanets. However, they suffer from
self-inconsistencies. A rotating, tidally deformed body does not in
fact possess multiple rotating tidal waves that create the
non-spherical equilibrium shape of a body. The properties of the tidal
bulge are due to rigidity, viscosity, structure and frequencies. ET
models are not much more than toy models for tidal evolution --
calculations from first principles would require three dimensions and include the
rheology of the interior and, for ocean-bearing worlds, a
3-dimensional model of currents, ocean floor topography and maps of
continental margins. For exoplanets, such a complicated model is not
available, nor is it necessarily warranted given the dearth of
observational constraints.

The ET frameworks permit fundamentally different assumptions regarding
the lag between the passage of the perturber and the passage of the
tidal bulge. This ambiguity has produced two well-developed models
that have reasonably reproduced observations in our Solar System, but
which can diverge significantly when applied to configurations with
different properties. One model assumes that the lag is a constant in
phase and is independent of frequency. In other words, regardless of
orbital and rotational frequencies, the phase between the perturber
and the tidal bulge remains constant. Following \cite{Greenberg09} I
will refer to this version as the ``constant-phase-lag" or CPL
model. At first glance, this model may seem to be the best choice,
given the body is expected to behave like a harmonic oscillator: In
order for the tidal waves to be linearly summed, the damping parameters must be
independent of frequency. However, for eccentric orbits, it may not be
possible for the phase lag to remain constant as the instantaneous
orbital angular frequency changes in accordance with Kepler's 2nd Law
\citep{ToumaWisdom94,EfroimskyMakarov13}. This paradox has led numerous
researchers to reject the CPL model, despite its relative success at
reproducing features in the Solar System,
\eg~\citep{MacDonald64,Hut81,GoldreichSoter66,Peale79}, as well as the
tidal circularization of close-in exoplanets \citep{Jackson08a}.

The second possibility for the lag is that the time interval between
the perturber's passage and the tidal bulge is constant. In this case,
as frequencies change, the angle between the bulge and the perturber
changes. I will call this version the ``constant-time-lag," or CTL,
model \citep{Mignard79,Hut81,Greenberg09}. The CPL and CTL models, while qualitatively different, reduce
to the same set of governing equations if a linear dependence between
phase lags and tidal frequencies is assumed.

In terms of planetary rotation rate, many of the timescales are set by
masses, radii, and semi-major axes. For typical main sequence stars,
and Mars- to Neptune-sized planets in the classic HZ of
\cite{Kasting93}, the timescales range from millions to trillions of
years, with the shortest timescales occurring for the largest planets
orbiting closest to the smallest stars. The CPL and CTL models predict
qualitatively similar behavior for the orbital evolution of close-in
exoplanets when rotational effects are ignored,
\eg~\citep{Jackson09,Levrard09,Barnes13,Barnes15_tides}.

The two tidal models are effectively indistinguishable in our Solar
System. Most tidally-interacting pairs of worlds are now evolving so
slowly that changes due to tidal evolution are rarely measurable. The
Earth-Moon system is by far the best measured, but interpretations are
hampered by the complexity of the binary's orbital history and the
dissipation of energy due to Earth's ever-changing continental
configuration \citep{Green17}. Nonetheless, the present state of the
Earth-Moon system provides crucial insight into the rotational
evolution of habitable exoplanets.

When the ET framework is limited to two bodies, it consists of 6
independent parameters: the semi-major axis $a$, the eccentricity $e$,
the two rotation rates $\Omega_i$, and two obliquities $\psi_i$, where
$i$ (=1,2) corresponds to one of the bodies. If the gravitational
gradient across a freely rotating body induces sufficient strain on
that body's interior to force movement, then frictional heating is
inevitable. The energy for this heating comes at the expense of the
orbit and/or rotation, and hence tidal friction decreases the
semi-major axis and rotation period. The friction will also prevent
the longest axis of the body to align exactly with the perturber. With
an asymmetry introduced, torques arise and open pathways for angular
momentum exchange. There are three reservoirs of angular momentum: the
orbit and the two rotations. ET models assume orbit-averaged torques,
which is a good approximation for the long-term evolution of the
system. The redistribution of angular momentum depends on the amount
of energy dissipated, and the heights and positions of the tidal
bulges relative to the line connecting the two centers of mass. The ET
model can therefore be seen as a mathematical construction that
couples energy transformation to conservation of angular momentum.

The tidal power and bulge properties depend on the composition and
microphysics of planetary and stellar interiors, which are very
difficult to measure in our Solar System, let alone in an external
planetary system. In ET theory, the coupling between energy
dissipation and the tidal bulge is therefore a central feature, and
are scaled by two parameters, the Love number of degree 2, $k_2$, and
a parameter that represents the lag between the line connecting the
two centers of mass and the direction of the tidal bulge. In the CPL
model, this parameter is the ``tidal quality factor" $Q$, and in CTL
it is the tidal time lag $\tau$. $k_2$ is the same in both models.

Although energy dissipation results in semi-major axis decay, angular
momentum exchange can lead to semi-major axis growth. This growth can occur
if enough rotational angular momentum can be transferred to the semi-major axis
to overcome its decay due to tidal heating. Earth and the Moon are in
this configuration now.

A tidally evolving system has three possible final configurations
\citep{Counselman73}: both bodies rotate synchronously with spin axes
parallel to the orbital axis and they revolve around each other on a
circular orbit; the two bodies merge; or the orbit expands
forever. The first case is called ``double synchronous rotation", and
is the only stable solution. In the latter case, encounters with other
massive bodies will likely prevent the unimpeded expansion of the orbit. As
the rotation rates decay, the less massive body will probably be the
first to reach the synchronous state, and this is the case for
habitable exoplanets of main sequence stars.

\subsection{The Constant Phase Lag Model}

In the 2nd order CPL model of tidal evolution, the angle between the line connecting the
centers of mass and the tidal bulge is constant. This approach is commonly
utilized in Solar System studies \citep[\eg][]{GoldreichSoter66,Greenberg09} and the
evolution is described by the following equations:

\begin{equation}\label{eq:e_cpl}
  \frac{\mathrm{d}e}{\mathrm{d}t} \ = \ - \frac{ae}{8 G M_1 M_2}
  \sum\limits_{i = 1}^2Z'_i \Bigg(2\varepsilon_{0,i} - \frac{49}{2}\varepsilon_{1,i} + \frac{1}{2}\varepsilon_{2,i} + 3\varepsilon_{5,i}\Bigg),
\end{equation}

\begin{equation}\label{eq:a_cpl}
  \frac{\mathrm{d}a}{\mathrm{d}t} \ = \ \frac{a^2}{4 G M_1 M_2}
  \sum\limits_{i = 1}^2 Z'_i  \ {\Bigg(} 4\varepsilon_{0,i} + e^2{\Big [} -20\varepsilon_{0,i} + \frac{1
47}{2}\varepsilon_{1,i} + \nonumber \frac{1}{2}\varepsilon_{2,i} - 3\varepsilon_{5,i} {\Big ]} -4\sin^2(\psi_i){\Big [}\varepsilon_{0,i}-\varepsilon_{8,i}{\Big ]}{\Bigg )}, 
\end{equation}

\begin{equation}\label{eq:o_cpl}
  \frac{\mathrm{d}\Omega_i}{\mathrm{d}t} \ = \ - \frac{Z'_i}{8 M_i r_{\mathrm{g},i}
^2 R_i^2 n} {\Bigg (}4\varepsilon_{0,i} + e^2{\Big [} -20\varepsilon_{0,i} + 49\varepsilon_{1,i} + \varepsilon_{2,i} {\Big ]} + \nonumber \ 2\sin^2(\psi_i) {\Big [} -
2\varepsilon_{0,i} + \varepsilon_{8,i} + \varepsilon_{9,i} {\Big ]} {\Bigg )}, \\  
\end{equation}

\noindent and 

\begin{equation}\label{eq:psi_cpl}
  \frac{\mathrm{d}\psi_i}{\mathrm{d}t} \ = \ \frac{Z'_i \sin(\psi_i)}{4 M_i r_{\mathrm{g},i}^2 R_i^2 n \Omega_i} {\Bigg (} {\Big [} 1-\xi_i {\Big ]}\varepsilon_{0,i} 
+ {\Big [} 1+\xi_i {\Big ]}{\Big \{}\varepsilon_{8,i}-\varepsilon_{9,i}{\Big \}} {\Bigg)} \ ,
\end{equation}

\noindent where $t$ is time, $G$ is Newton's gravitational constant, $M_i$ are the two
masses, $R_i$ are the two radii, and $n$ is the mean
motion. The above equations are mean variations of the orbital
elements, averaged over the orbital the period, and are only valid to
second order in $e$ and $\psi$. The quantity $Z'_i$ is

\begin{equation}\label{eq:Zp}
Z'_i \equiv 3 G^2 k_{2,i} M_j^2 (M_i+M_j) \frac{R_i^5}{a^9} \ \frac{1}{n Q_i} \ ,
\end{equation}

\noindent where $k_{2,i}$ are the Love numbers of order 2, and $Q_i$ are the ``tidal quality factors.'' The parameter $\xi_i$ is

\begin{equation}\label{eq:chi}
\xi_i \equiv \frac{r_{\mathrm{g},i}^2 R_i^2 \Omega_i a n }{ G M_j},
\end{equation}

\noindent where $i$ and $j$ refer to the two bodies, and $r_g$ is the ``radius of gyration,'' \ie the moment of inertia is $M(r_gR)^2$. The signs of the phase lags are

\begin{equation}\label{eq:epsilon}
\begin{array}{l}
\varepsilon_{0,i} = \Sigma(2 \Omega_i - 2 n)\\
\varepsilon_{1,i} = \Sigma(2 \Omega_i - 3 n)\\
\varepsilon_{2,i} = \Sigma(2 \Omega_i - n)\\
\varepsilon_{5,i} = \Sigma(n)\\
\varepsilon_{8,i} = \Sigma(\Omega_i - 2 n)\\
\varepsilon_{9,i} = \Sigma(\Omega_i) \ ,\\
\end{array}
\end{equation}

\noindent with $\Sigma(x)$ the sign of any physical quantity $x$, \ie
$\Sigma(x)~=~+1, -1$ or 0. 

The CPL model described above only
permits 4 ``tidal waves'', and hence does not allow this continuum,
only spin to orbit frequency ratios of 3:2 and
1:1. Essentially, for eccentricities below $\sqrt{1/19}$, the torque on
the rotation by the tidal waves is insufficient to change the rotation
rate, but above it, the torque is only strong enough to increase the
rotation rate to $1.5n$. In other words, these rotation rates are the
only two resolved by the truncated infinite series used in the CPL
framework, and the equilibrium rotation period is
\begin{equation}\label{eq:p_eq_cpl}
P^{CPL}_{eq} = 
\begin{dcases}
  P, & e < \sqrt{1/19}\\
  \frac{2P}{3}, & e~\le~\sqrt{1/19},
\end{dcases}
\end{equation}
where $P$ is the orbital period. \citep{Goldreich66} suggested that the equilibrium rotation period, \ie the rotation period of a ``tidally locked'' body, is
\begin{equation}\label{eq:p_eq_g66}
P_{eq}^{G66} = \frac{P}{1 + 9.5e^2}.
\end{equation}
\citep{MurrayDermott99} presented a derivation of this expression, which
assumes $\sin(\psi) = 0$, and predicts the rotation rate may take
a continuum of values.

\cite{Barnes13} suggested the CPL model should be
implemented differently depending on the problem. When modeling the
evolution of a system, one should use the discrete spin values for
self-consistency, \ie as an initial condition, or if forcing the spin
to remain tide-locked. However, if calculating the equilibrium spin
period separately, the continuous value of Eq.~(\ref{eq:p_eq_cpl})
should be used. I refer to these rotational states as ``discrete'' and
``continuous'' and will use the former throughout this study.

Note in Eq.~(\ref{eq:e_cpl}) that $e$ can grow if 
\begin{equation}{\label{eq:eccgrowth_cpl}}
\sum\limits_{i = 1}^2Z_i\Bigg(2\varepsilon_{0,i} - \frac{49}{2}\varepsilon_{1,i} + \frac{1}{2}\varepsilon_{2,i} + 3\varepsilon_{5,i}\Bigg) < 0,
\end{equation}
which depends on the orbital
and rotational frequencies, as well as the physical properties of the
two bodies. Eccentricity growth by tidal effects has been considered
for brown dwarfs \citep{Heller10}, but has not been examined for
potentially habitable exoplanets. If $e$ grows, then by
Eq.~(\ref{eq:p_eq_cpl}), we expect a planet's equilibrium rotational period to
decrease, \ie it will never be synchronized. Thus, synchronization is
not necessarily the end state of the tidal evolution of a planet's
spin, if the equilibrium tide
models are approximately valid at high eccentricity.

\subsection{The Constant Time Lag Model}

The CTL model assumes that the time interval between the passage of
the perturber and the tidal bulge is constant. This assumption allows
the tidal response to be continuous over a wide range of frequencies,
unlike the CPL model. But, if the phase lag is a function of the
forcing frequency, then the system is no longer analogous to a damped
driven harmonic oscillator. Therefore, this model should only be used
over a narrow range of frequencies, see \citep{Greenberg09}. However,
this model's use is widespread, especially at high $e$, so I use it to
evaluate tidal effects as well.

Here I use the model derived by \citep{Leconte10}, with the nomenclature of \cite{Heller11}. The evolution is described by the following equations:

\begin{equation} \label{eq:e_ctl}
  \frac{\mathrm{d}e}{\mathrm{d}t} \ = \ \frac{11 ae}{2 G M_1 M_2}
  \sum\limits_{i = 1}^2Z_i \Bigg(\cos(\psi_i) \frac{f_4(e)}{\beta^{10}(e)}  \frac{\Omega_i}{n} -\frac{18}{11} \frac{f_3(e)}{\beta^{13}(e)}\Bigg),
\end{equation}

\begin{equation}\label{eq:a_ctl}
  \frac{\mathrm{d}a}{\mathrm{d}t} \ = \  \frac{2 a^2}{G M_1 M_2}
  \sum\limits_{i = 1}^2 Z_i \Bigg(\cos(\psi_i) \frac{f_2(e)}{\beta^{12}(e)} \frac{\Omega_i}{n} - \frac{f_1(e)}{\beta^{15}(e)}\Bigg),
\end{equation}

\begin{equation}\label{eq:o_ctl}
  \frac{\mathrm{d}\Omega_i}{\mathrm{d}t} \ = \ \frac{Z_i}{2 M_i r_{\mathrm{g},i}^2 
R_i^2 n} \Bigg( 2 \cos(\psi_i) \frac{f_2(e)}{\beta^{12}(e)} - \left[ 1+\cos^2(\psi)
 \right] \frac{f_5(e)}{\beta^9(e)} 
\frac{\Omega_i}{n} \Bigg),  
\end{equation}

\noindent and

\begin{equation}\label{eq:psi_ctl}
  \frac{\mathrm{d}\psi_i}{\mathrm{d}t} \ = \ \frac{Z_i \sin(\psi_i)}{2 M_i r_{\mathrm{g},i}^2 R_i^2 n \Omega_i}\left( \left[ \cos(\psi_i) - \frac{\xi_i}{ \beta} \right] \frac{f_5(e)}{\beta^9(e)} \frac{\Omega_i}{n} - 2 \frac{f_2(e)}{\beta^{12}(e)} \right).
\end{equation}

\noindent where

\begin{equation}\label{eq:Z}
 Z_i \equiv 3 G^2 k_{2,i} M_j^2 (M_i+M_j) \frac{R_i^5}{a^9} \ \tau_i \ ,
\end{equation}

\noindent and 

\begin{equation}\label{eq:f_e}
\begin{array}{l}
\beta(e) = \sqrt{1-e^2},\\
f_1(e) = 1 + \frac{31}{2} e^2 + \frac{255}{8} e^4 + \frac{185}{16} e^6 + \frac{25}{
64} e^8,\\
f_2(e) = 1 + \frac{15}{2} e^2 + \frac{45}{8} e^4 \ \ + \frac{5}{16} e^6,\\
f_3(e) = 1 + \frac{15}{4} e^2 + \frac{15}{8} e^4 \ \ + \frac{5}{64} e^6,\\
f_4(e) = 1 + \frac{3}{2} e^2 \ \ + \frac{1}{8} e^4,\\
f_5(e) = 1 + 3 e^2 \ \ \ + \frac{3}{8} e^4.
\end{array}
\end{equation}

\begin{figure} 
\begin{center}
\includegraphics[width=0.9\textwidth]{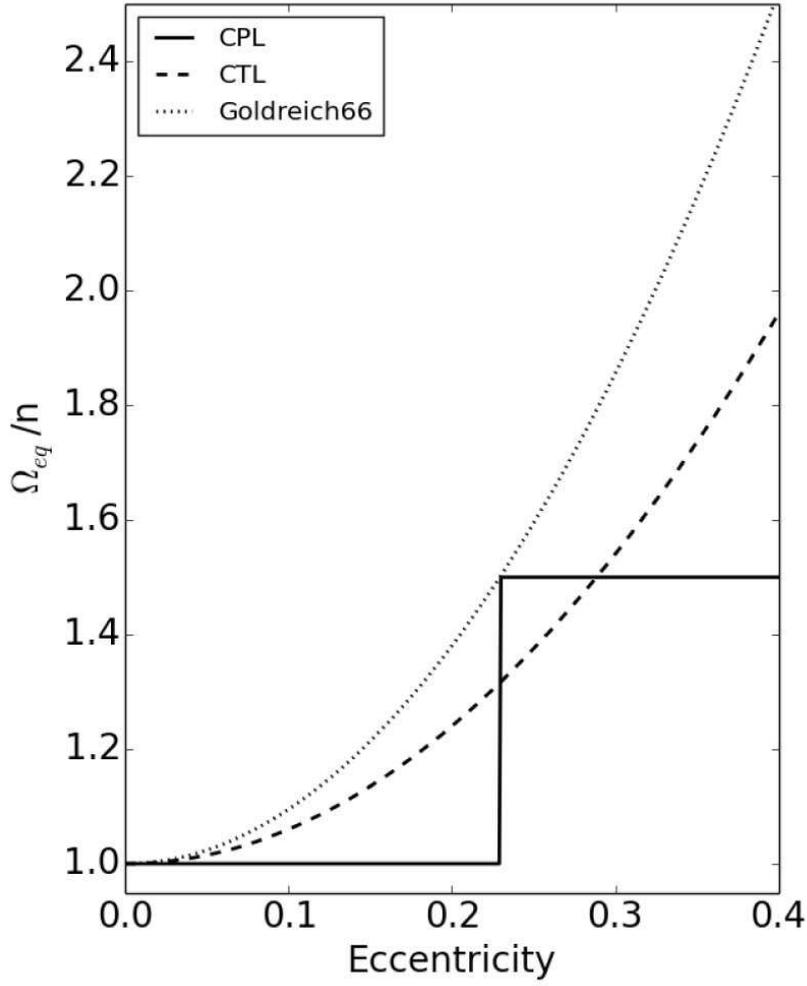}
\end{center}
\caption{Equilibrium spin frequency of a tidally locked exoplanet as a function of eccentricity for the  CPL model (solid curve), the CTL model (dashed), and \cite{Goldreich66}'s model (dotted).}
\label{fig:eqspin}
\end{figure}

It can also be shown that the equilibrium rotation period is
\begin{equation}\label{eq:p_eq_ctl_obl}
P_{eq}^{CTL}(e,\psi) = P\frac{\beta^3f_5(e)(1 + \cos^2\psi)}{2f_2(e)\cos\psi},
\end{equation}
which for low $e$ and $\psi = 0$ reduces to
\begin{equation}\label{eq:p_eq_ctl}
P_{eq}^{CTL} = \frac{P}{1 + 6e^2}.
\end{equation}
Fig.~\ref{fig:eqspin} shows the predicted ratio of the equilibrium
rotational frequency $\Omega_{eq}$ to $n$ as a function of $e$ for the
CPL model (solid curve), \cite{Goldreich66}'s model (dotted) and the
CTL model (dashed). All models predict that as $e$ increases, the
rotational frequency of a tidally locked planet will grow, and hence
planets found on eccentric orbits may rotate super-synchronously.

As in the CPL model, the CTL model also predicts configurations that
lead to eccentricity growth, which could prevent rotational
synchronization. If
\begin{equation}\label{eq:eccgrowth_ctl}
 \sum\limits_{i = 1}^2Z_i \Bigg(\cos(\psi_i) \frac{f_4(e)}{\beta^{10}(e)}  \frac{\Omega_i}{n} -\frac{18}{11} \frac{f_3(e)}{\beta^{13}(e)}\Bigg) > 0
 \end{equation}
 then the orbital eccentricity will grow.

\subsection{Numerical Methods (\eqtide)}

To survey the range of rotational evolution of habitable exoplanets,
I performed thousands of numerical simulations of individual
star-planet pairs and tracked their orbital and rotational states. To
compute the models described in the previous two subsections I used
the software package \eqtide\footnote{Publicly available at
  https://github.com/RoryBarnes/EqTide.}. This code, written in C,
is fast, user-friendly, and has a simple switch between the CTL and
CPL models. The governing equations are calculated with a 4th order
Runge-Kutta integrator (although Euler's method works surprisingly
well), with an adaptive timestep determined by the most rapidly
evolving parameter. This software package and its variants have been
used on a wide range of binary systems, including exoplanets
\citep{Barnes13,Barnes15_tides}, binary stars
\citep{GomezMaqueoChew12,GomezMaqueoChew14}, brown dwarfs
\citep{Fleming12,Ma13}, and exomoons \citep{HellerBarnes13}.

For the simulations presented below, I used the Runge-Kutta method
with a timestep that was 1\% of the shortest dynamical time
($x/(dx/dt)$, where $x$ is one of the 6 independent variables), and
assumed a planet became tidally locked if its rotation period reached
1\% of the equilibrium rotation period. If a planet becomes tidally
locked, its rotation period is forced to equal the equilibrium period (Eq.~[\ref{eq:p_eq_cpl}] or Eq.~[\ref{eq:p_eq_ctl}])
for the remainder of the integration. Note that the obliquity,
eccentricity and mean motion can continue to evolve and so the spin
period can also.

\subsection{What is the Tidal Response?}

The ET models described above have two free parameters: $k_2$ and
either $Q$ or $\tau$. The former can only have values between 0 and
1.5, but the latter can span orders of magnitude. As is well known, the current estimates
for $Q$ and $\tau$ of Earth, 12 and 640~s, respectively, predict that
the Moon has only been in orbit for 1--2~Gyr
\citep{MacDonald64,ToumaWisdom94}, far less than its actual age of 4.5 Gyr. This contradiction has led
many researchers to assume that the historical averages of Earth's
tidal $Q$ and $\tau$ have been about 10 times different. Indeed,
\cite{Kasting93} assumed $Q=100$ when they calculated their tidal lock
radius. Previous researchers have argued that such discrepancies are
justified because tidal dissipation in the oceans is a complex
function of continental positions and in the past different
arrangements could have allowed for weaker dissipation. A recent study
of ocean dissipation over the past 250 Myr found that the modern Earth
is about twice as dissipative as compared to the historical average
\citep{Green17}. Another possibility is that the continents grow with
time
\citep{Dhuime12} resulting in a larger fraction of the surface
consisting of highly dissipative straits and shallow seas. Whether
these suppositions are true remains to be seen, but clearly the ET
models fail when using modern values of the tidal dissipation. On the
other hand, if $Q$ ($\tau$) has been shrinking (growing) then we might
expect it to continue to do so, and planets older than Earth may
be more dissipative.

However, one also needs to bear in mind the complexities of the
Earth-Moon system's presence in the Solar System. \cite{Cuk07} showed
that the Moon's orbit and obliquity have been perturbed by passages
through secular resonances with Venus and Jupiter as the Moon's orbit
expanded. During these perturbations, the eccentricity of the Moon
could have reached 0.2 and hence the simple picture invoked by classic
studies of the two-body Earth-Moon system may be invalid (and that
there may have been epochs in Earth history during which the Moon rotated
super-synchronously). Moreover, the early evolution of the Earth-Moon
system may have been affected by an evection resonance involving the
Earth's orbit about the Sun and by energy dissipation inside the solid
Earth \citep{CukStewart12,Zahnle15}. Given these complications, it is
unclear that the modern dissipation of Earth is not representative of
its historical average, nor is it obvious that other ocean-bearing
exoplanets should tend to have less dissipation than Earth.

\begin{figure} 
\begin{center}
\includegraphics[width=0.8\textwidth]{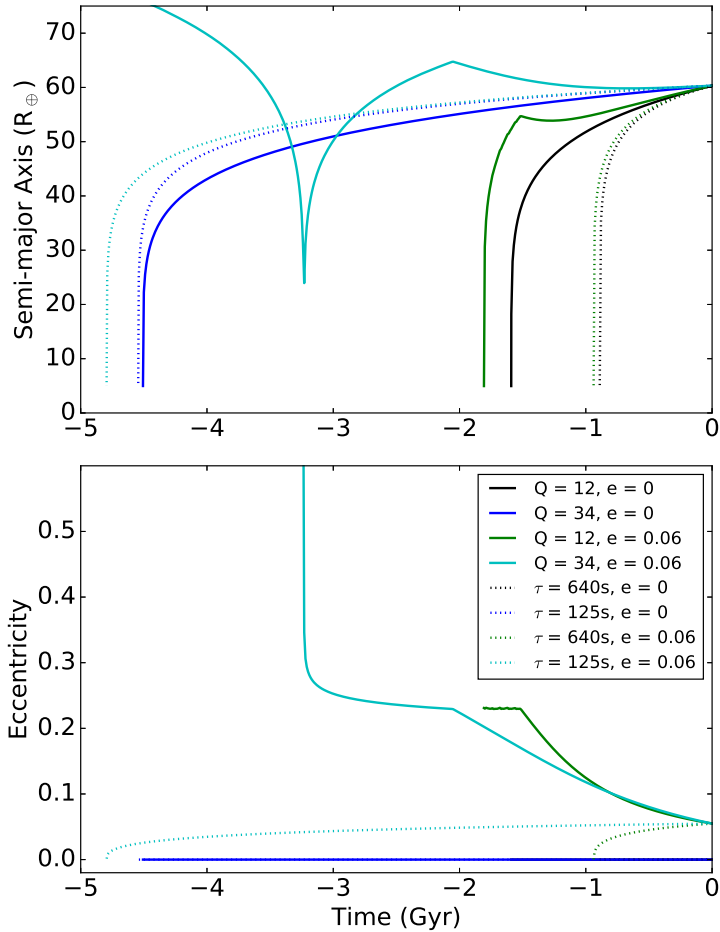}
\end{center}
\caption{Tidal evolution of the semi-major axis (top) and eccentricity (bottom) of the Earth-Moon
  system for different assumptions. Solid curves correspond to the CPL
  model, dotted to CTL. The black curves assume the modern value of
  Earth's tidal response, but the eccentricity and 2 obliquities have
  been set to 0. These curves reproduce the classic result that the
  Moon was at Earth's surface 1--2 Gyr ago. The blue curves show the
  history if the $Q$ and $\tau$ values are chosen so that the Moon's
  orbit began expanding 4.5 Gyr ago. The green curves use the modern
  tidal response and the current values of $e$ and $\psi$. Cyan curves
  are similar, but use the calibrated values of $Q$ and $\tau$.}
\label{fig:earthmoon}
\end{figure}

In Fig.~\ref{fig:earthmoon} the semi-major axis and eccentricity history of the
Earth-Moon system is shown for the two ET models with different
choices for the tidal response and different approximations. For the
simple case of a circular orbit and no obliquities, I recover the
classic results. However, dissipation rates that are 10 times less
than the current rate, as advocated by \cite{Kasting93}, do not
accurately calibrate the models either. Instead I find rates that are
3--5 times weaker place the Moon at Earth's surface 4.5 Gyr ago
($Q=34.5$ and $\tau=125$~s). When the assumption that $e~=~\sin\psi~=~0$
is relaxed and the modern values are used, the CTL model predicts
slightly different behavior, but the CPL model is qualitatively
different and actually predicts large values of $a$ and $e$ in the
past. This history is strongly influenced by the current values of $e$
and $\psi$, which were modified by the secular resonance
crossings \citep{Cuk07}.

In summary, the ET models predict reasonable histories for the
Earth-Moon system, but the history is sufficiently complicated that we
cannot unequivocally state that the modern Earth's $Q$ and $\tau$
values should be discarded when considering habitable exoplanets. It
is also currently impossible to know how different continents and
seafloor topography affect the tidal response of habitable
exoplanets. In light of these uncertainties, I will use the modern
Earth's values of $Q$ and $\tau$ in the simulations of the tidal
evolution of habitable exoplanets. In general the timescales scale
linearly with these parameters, so if one prefers different choices,
it is relatively straight-forward to estimate the evolution based on
the plots provided below, or to directly calculate it with \eqtide.

\subsection{Initial Conditions}

Every trial run requires specific physical and orbital properties,
some of which can be constrained by models, but others cannot.  A
star's mass is roughly constant over its main sequence lifetime, but
its radius and luminosity can change significantly over time. I fit
the 5 Gyr mass-radius-luminosity relation from \cite{Baraffe15} to a
third order polynomial using Levenberg-Marquardt minimization
\citep{Press92} and found
\begin{equation}\label{eq:starmr}
\frac{R_*}{R_\odot} = 0.003269 + 1.304\frac{M_*}{M_\odot} - 1.312\frac{M_*}{M_\odot}^2 + 1.055\frac{M_*}{M_\odot}^3,
\end{equation}
and
\begin{equation}\label{eq:starml}
\log\Big(\frac{L_*}{L_\odot}\Big) = -0.0494 + 6.65X + 8.73X^2 + 5.208X^3,
\end{equation}
where $X = \log (M_*/M_\odot)$. The star's $k_2$, $Q$, and $\tau$
values are set to 0.5, $10^6$, and 0.01~s for all cases, similar to
previous studies
\citep{Rasio96,Jackson08a,Matsumura10,Barnes15_tides}. For the CPL
model, the initial values of the phase lags are determined by the
initial rotational and orbital frequencies as described in
Eq.~(\ref{eq:epsilon}).

For the planet, I used the mass-radius relationship of \cite{Sotin07}
for planets with Earth-like composition:
\begin{equation}
\label{eq:rterrsotin07}
\frac{R_p}{\rearth} = \left\{ \begin{array}{rl}
 \Big(\frac{M_p}{\mearth}\Big)^{0.306}, &\mbox{ $10^{-2}~\mearth < M_p < \mearth$} \\
 \Big(\frac{M_p}{\mearth}\Big)^{0.274}, &\mbox{ $\mearth < M_p < 10~\mearth$}.
       \end{array} \right.
\end{equation}
The planet's $k_2$, $Q$, and $\tau$ values are set to 0.3, 12 and
640~s \citep{Lambeck77,Williams78,Yoder95}.

Fixing each of the physical parameters as described does not represent
a comprehensive study of the problem, as planets and stars of a given
mass can have a wide range of radii and tidal response. However, as
shown in the next section, using these standard parameters and
allowing the initial rotation period to vary admits a wide enough
range of possibilities that there is no need to explore the other
parameters at this time.

I also calculate the predicted tidal evolution of planets found by
\kepler, and projected for \tess. For these simulations, I used both
the CPL and CTL model, and three sets of initial conditions: ($P_0,
\psi_0, Q, \tau$) = (10 d, 0, 12, 640 s), (1 d, 23.5$^\circ$, 34, 125
s), and (8 hr, 60$^\circ$, 100, 64 s), where $P_0$ is the initial
rotation period. I will refer to these options as ``short'',
``Earth-like'', and ``long'' timescales for $T_{lock}$, respectively.

\section{Results}\label{sec:results}

This section considers the tidal evolution of ocean-bearing exoplanets
orbiting various stars and with different, but reasonable, initial and
physical conditions with the goal of mapping out regions of parameter
space that lead to synchronous rotation on Gyr timescales. First, I
consider the tidal evolution of Kepler-22 b, a $2.3~\rearth$ planet
orbiting a G5V star near the inner edge of its HZ with no known
companions~\citep{Borucki12}. This planet is probably too large to be
terrestrial \citep{Rogers15}, but is similar to others worlds that may
be~\citep{Dumusque14,Espinoza16}, and so is potentially habitable
\citep{Barnes15_hite}. Next I examine the tidal evolution of the
recently-discovered planet Proxima Centauri b
\citep{AngladaEscude16}. Then I consider the potentially habitable
planets discovered by the {\it Kepler} spacecraft that do not have
planetary companions. Then I survey a broad range of parameter space
and show that habitable exoplanets of G dwarf stars with an initially
slow rotation period and low obliquity can become tidally locked within
1~Gyr. Next I explore the coupled orbital/rotational evolution of
planets orbiting very small stars ($\lsim~0.1~\msun$) and find that
they are likely to be synchronous rotators. Finally, I compute the
rotational evolution of the projected exoplanet yield from {\it TESS}
\citep{Sullivan15} and find that nearly every potentially habitable
planet the mission will discover will be tidally locked.

\subsection{Kepler-22 b}
\begin{figure} 
\includegraphics[width=\textwidth]{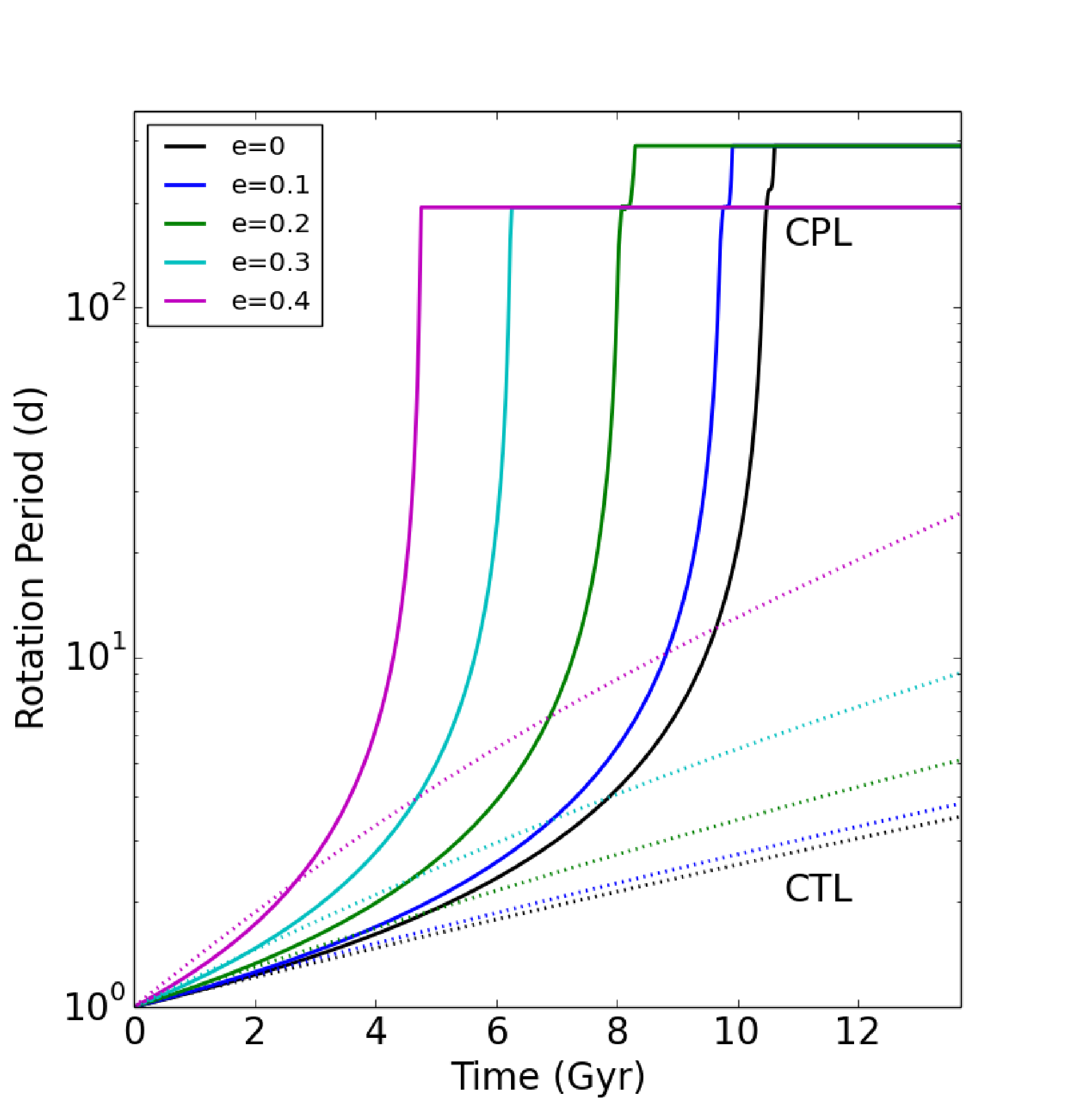}
\caption{Evolution of the rotation period of Kepler-22 b with
  different orbital eccentricities as shown in the legend. Solid lines
  assume the CPL model, dashed CTL. The evolution of semi-major axis and eccentricity is negligible.}
\label{fig:k22b}
\end{figure}

Fig.~\ref{fig:k22b} shows plausible spin period evolutions of
Kepler-22 b assuming a mass of $23~\mearth$, an initial spin period of
1 day, and an initial obliquity of 23.5$^\circ$. The mass of the host star is 0.97~$\msun$ and the initial semi-major axis is 0.849~AU. The solid curves show
results from the CPL model, dashed CTL. The different line colors
represent different initial eccentricity, which is not
well-constrained at this time. The age of the system is not known, but
could be 10~Gyr~\citep{Borucki12}. For such an age, the CPL model
predicts that the planet will have spun down into a synchronous state
for $e~\lsim~0.2$ and into a 3:2 spin-orbit frequency ratio for larger
values. On the other hand, the CTL model predicts significantly less
evolution. This system demonstrates that the CPL model predicts that
tidal braking can be important in the HZs of G and K dwarfs, not just M
dwarfs as is usually assumed. 

\subsection{Proxima Centauri b}
Recently, \citep{AngladaEscude16} announced the discovery of a
$\gsim1.27~\mearth$ planet orbiting in the HZ of Proxima Centauri with a
period of 11.2 days and an inferred semi-major axis of 0.0485~AU. The
orbital eccentricity could only be constrained to be less than 0.35,
and since the discovery was made via radial velocity data, the radius
and actual mass are not known. \cite{Barnes16_prox} considered its
tidal evolution in the CPL framework and found that the rotational
frequency is synchronized in less than $10^6$ years and that the
orbital eccentricity is damped with a characteristic timescale of
1~Gyr. In this subsection I expand on those results and also consider
evolution in the CTL model.

\begin{figure} 
\includegraphics[width=\textwidth]{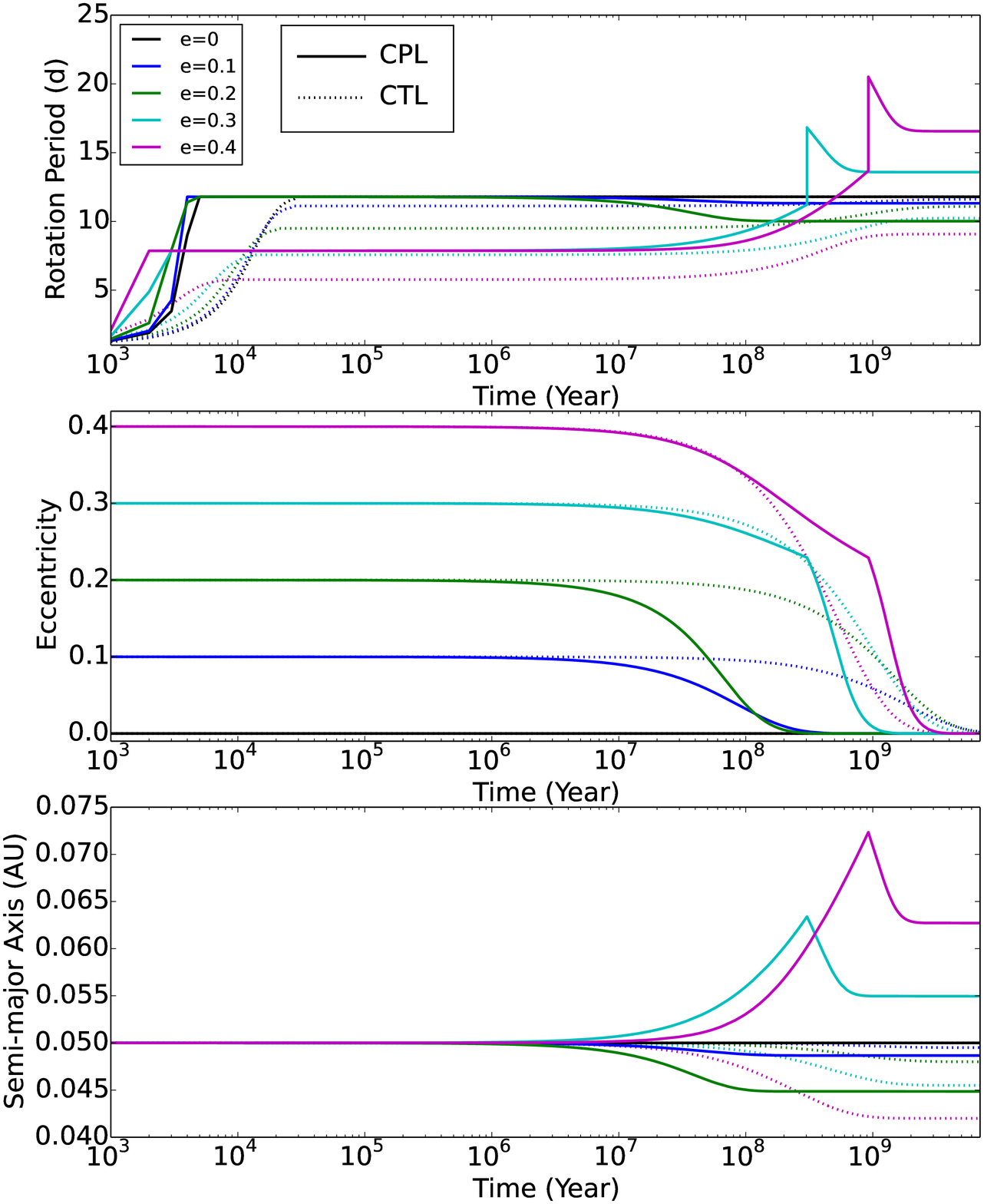}
\caption{Evolution of Proxima Centauri b due to tides. Solid lines
  assume the CPL model, dashed CTL. {\it Top:} Rotation period. {\it Middle:} Eccentricity. {\it Bottom:} Semi-major axis.}
\label{fig:proxima}
\end{figure}

In Fig.~\ref{fig:proxima} the evolution of rotation period, $e$
and $a$ are shown for a planet with Proxima b's minimum mass, a radius
of $1.07~\rearth$, an initial rotation period of 1 day, an initial
obliquity of 23.5$^\circ$, and an initial semi-major axis of
0.05~AU. The different linestyles correspond to different initial
eccentricities and different tidal models.

The CPL model predicts the rotational frequency becomes locked in less
than $10^4$ years, and that for $e~>~0.23$ planet b settles into a 3:2
spin-orbit frequency ratio that leads to semi-major axis growth. As
$a$ increases, so does the rotation period until
$e~\approx~0.23$ at which point the discrete nature of the CPL
solution instantaneously forces the synchronous state. At this point,
the torques are reversed and $a$ begins to decrease, ultimately
settling into orbits that are significantly larger than the initial
orbits.

The CTL model predicts the tidal locking time to be less than $10^5$
years, and that for $e~>~0$ the planet will reach
non-synchronous ``pseudo-equilibrium'' periods that are a function of
$e$. The CTL model does not suffer from discontinuities and predicts
longer times for $e$ to damp. Thus, if Proxima b ever reached a high
eccentricity state, the two ET models predict qualitatively different
types of evolution.

\subsection{The {\it Kepler} Sample}

\begin{figure} 
\includegraphics[width=\textwidth]{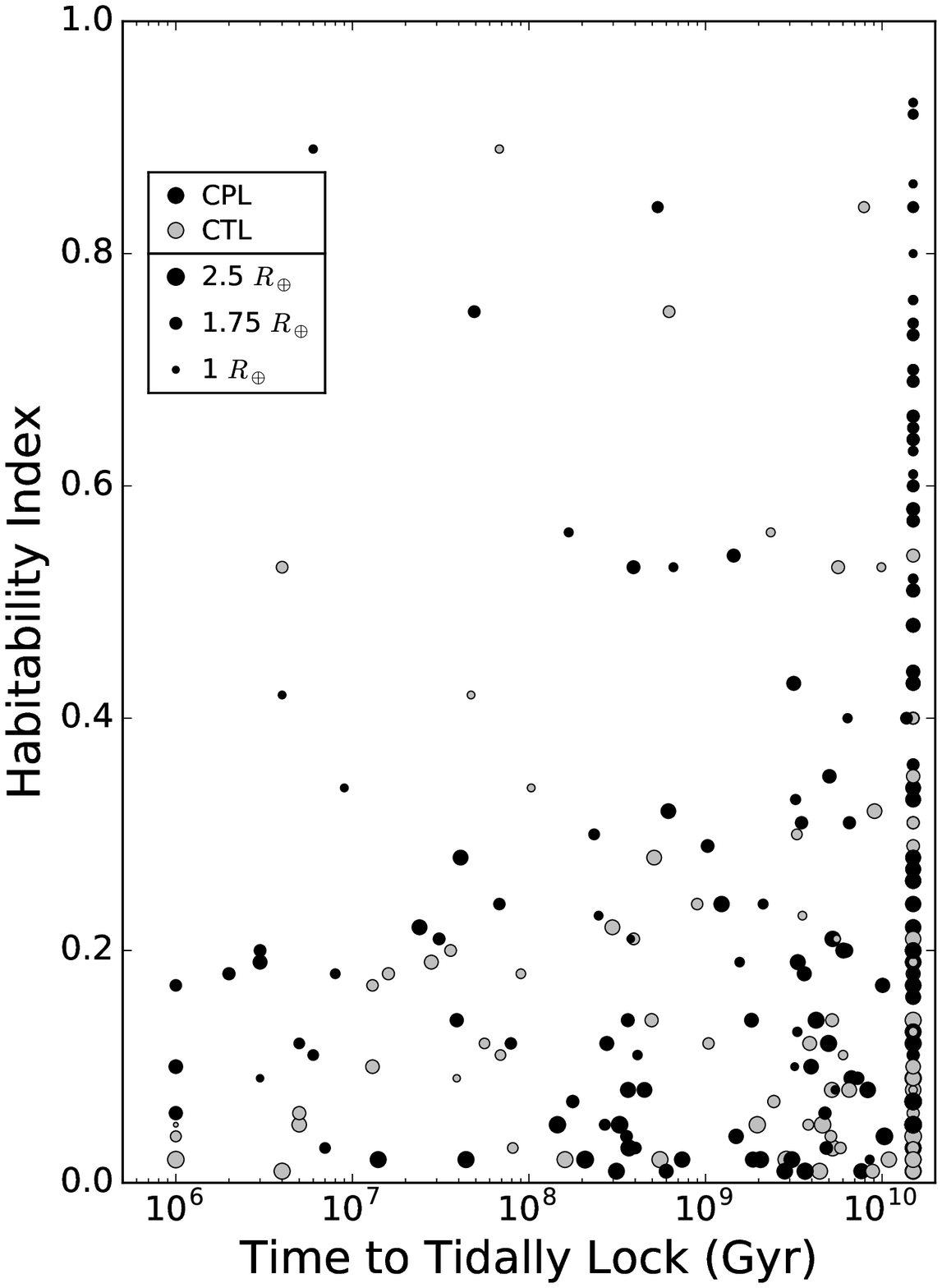}
\caption{Comparison of the time to tidally lock and the habitability index for transiting exoplanets. The dot size is proportional to planet mass, which is part of the habitability index, and the color corresponds to the equilibrium tide model. Planet candidates that did not tidally lock are placed at 15 Gyr.}
\label{fig:kepler}
\end{figure}

The previous examples are illustrative of the tidal evolution process
for potentially habitable planets, but are not comprehensive. In this
subsection, the analysis is expanded to the {\it Kepler} candidate
planets for which no other planets are known in the
system. Fig.~\ref{fig:kepler} shows the time for {\it Kepler}
candidates to tidally lock assuming they formed with 1 day rotation
periods and obliquities of $23.5^\circ$. Also shown is each
candidate's ``habitability index for transiting exoplanets'' (HITE)
\citep{Barnes15_hite}, which is an estimate of the likelihood that the
energy received at the top of the planet's atmosphere could permit
liquid surface water and is inversely proportional to the planet's
radius, as larger planets are more likely to be gaseous and
uninhabitable. For each planet candidate in \cite{Batalha13}, I
calculated the tidal evolution for 6 assumptions.

The column of planets at 15 Gyr is those candidates that did not tidally
lock within 15 Gyr. About half of the isolated and potentially
habitable planets discovered by {\it Kepler} could be tidally
locked. Table 1 (available in the Online Supplement) lists the properties of the planet candidates shown in
Fig.~\ref{fig:kepler}. The previously undefined symbols are: $P_{orb}$
= orbital period; $T_{CPL}^{short}$ = CPL model, ``short'' assumptions
(see $\S$~\ref{sec:methods}.5); $T_{CPL}^{\oplus}$ = CPL model,
Earth-like assumptions; $T_{CPL}^{long}$ = CPL model, ``long''
assumptions; $T_{CTL}^{short}$ = CTL model, short assumptions;
$T_{CTL}^{\oplus}$ = CTL model, Earth-like assumptions;
$T_{CTL}^{long}$ = CTL model, long assumptions.

\subsection{Parameter Space Survey}

\begin{figure} 
\begin{center}
\begin{tabular}{cc}
\includegraphics[width=0.4\textwidth]{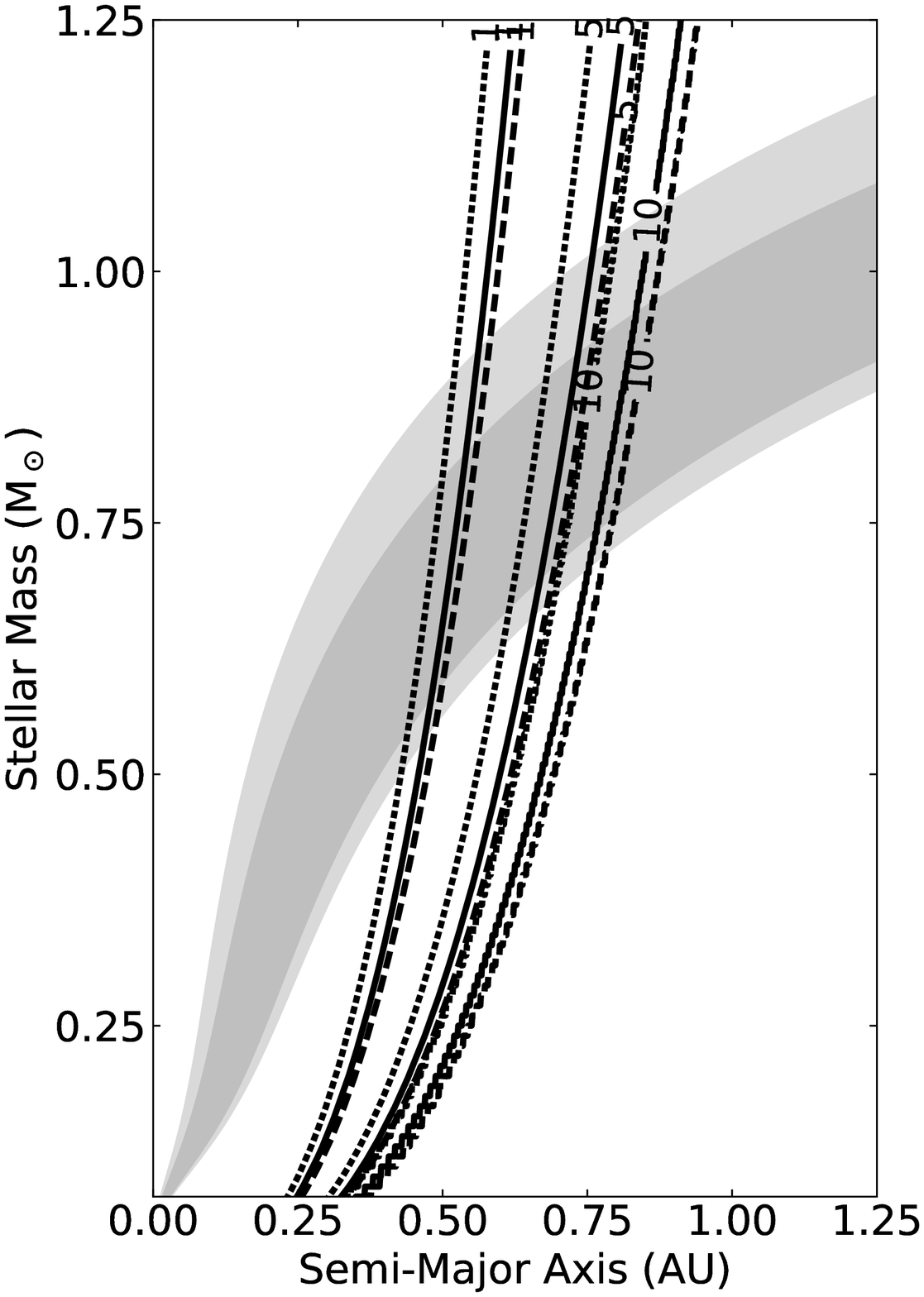} &
\includegraphics[width=0.4\textwidth]{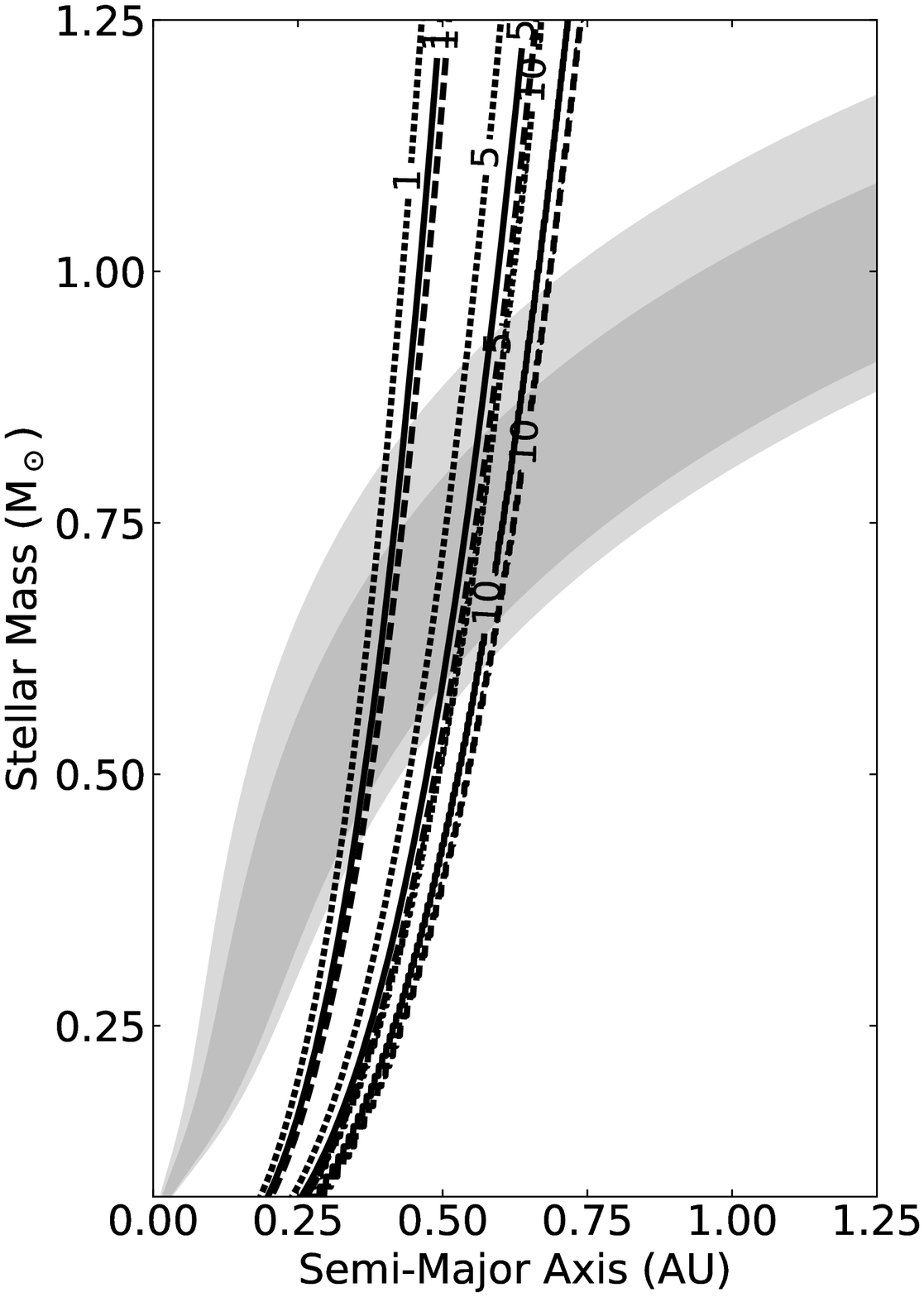}\\
\includegraphics[width=0.4\textwidth]{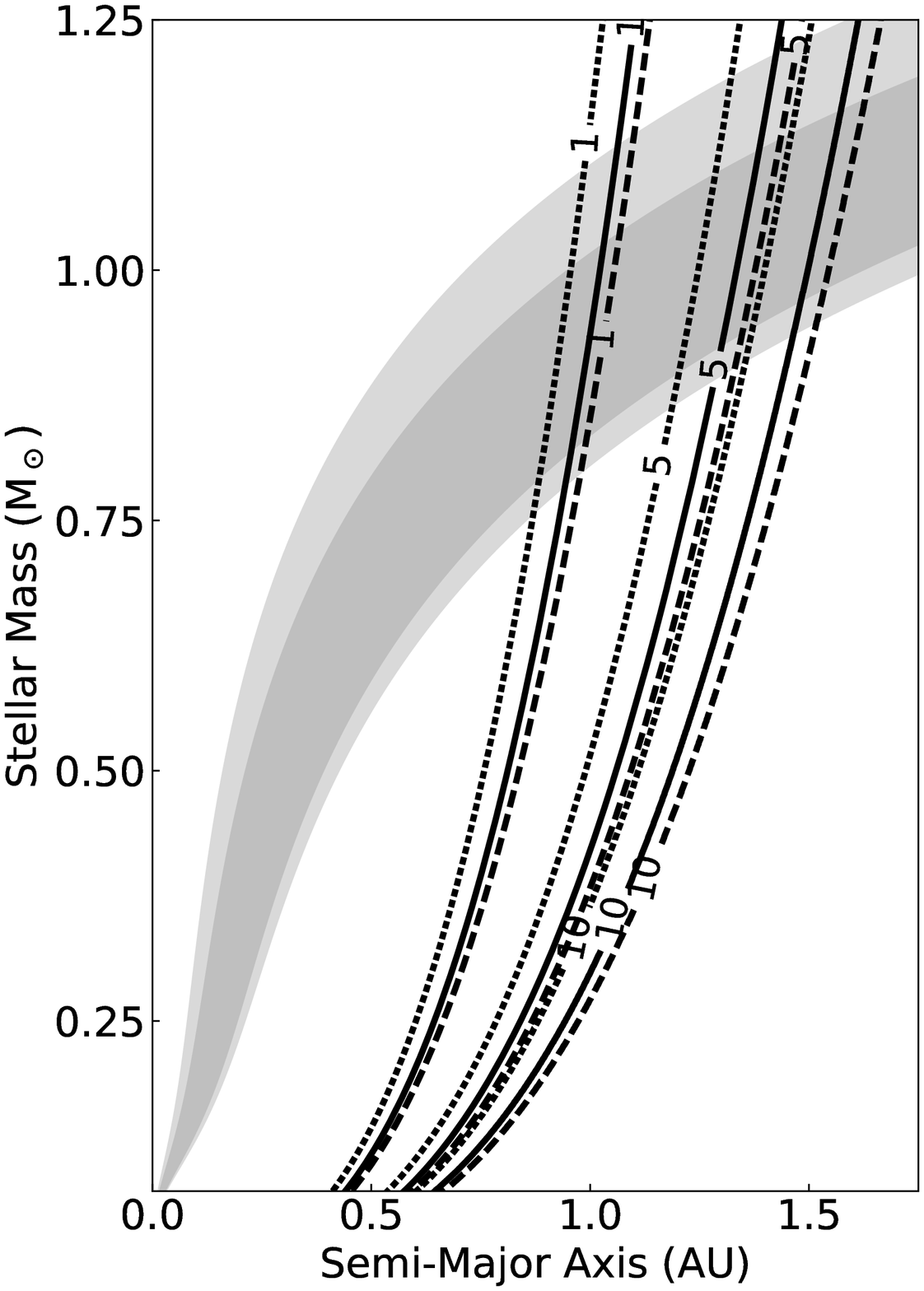} &
\includegraphics[width=0.4\textwidth]{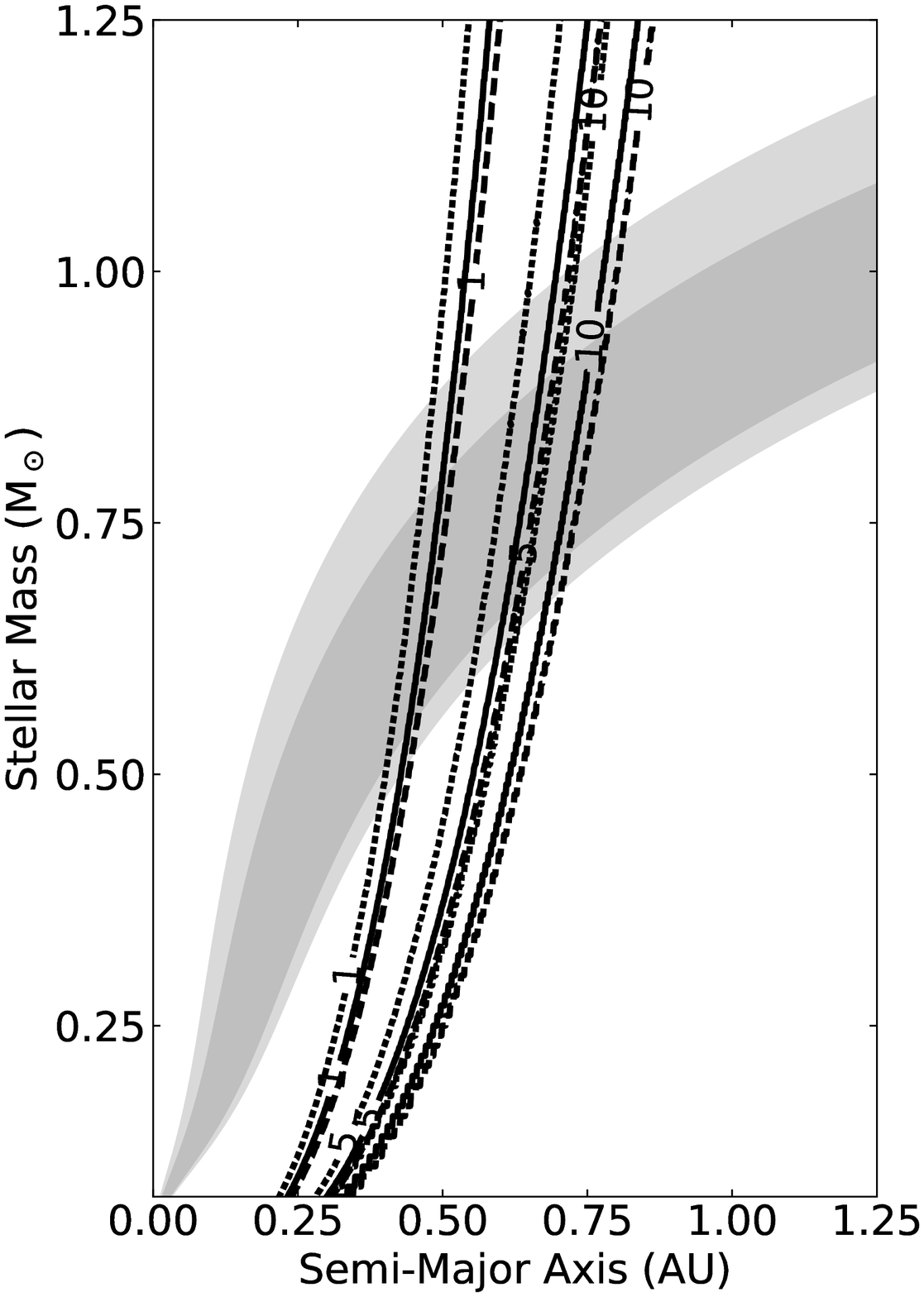}
\end{tabular}
\end{center}
\caption{Timescale for planets on circular orbits to rotate
  synchronously. The left panels assume the CPL model; right CTL. The
  top row assumes an initial rotation period of 8~hr and an obliquity
  of $60^\circ$. The bottom row assumes an initial rotation period of
  10 days and an obliquity of 0. The HZ is shown by the gray shading,
  with the darker gray corresponding to the conservative HZ, and
  lighter grey to the optimistic HZ \citep{Kopparapu13}. Note the
  different scale on the bottom left plot.}
\label{fig:survey}
\end{figure}

Next, I consider a broader range of parameter space to explore the
limits of the timescale to synchronize habitable exoplanets on
circular orbits. Fig.~\ref{fig:survey} shows the results of four
parameter sweeps. The left column is the CPL framework, right is
CTL. The top row is the more stable case, a planet with an initial
spin period of 8 hr and an initial obliquity of $60^\circ$. The bottom
row shows the more unstable case, a planet with an initial spin period
of 10 days, and an initial obliquity of 0. The former has
higher initial rotational angular momentum and its direction is
misaligned with the orbital angular momentum, and hence it takes
longer to equilibrate into the synchronous state. The latter is a
plausible initial condition from simulations of terrestrial planet
formation \citep{KokuboIda07,MiguelBrunini10}. The HZ of
\cite{Kopparapu13} is shown as shading with the lighter gray
representing empirical or optimistic limits, and the darker gray
representing theoretical or conservative limits.

This figure demonstrates that for early M through G dwarfs, the
possibility of rotational synchronization of habitable exoplanets on
circular orbits depends on the initial conditions and the tidal
response. Slower spin periods with aligned rotational and orbital axes
are more likely to lead to tidal synchronization within 1
Gyr. Earth-mass planets in the HZ of the entire G spectral class can
be synchronized if the CPL model is correct, but if CTL is, then
planets orbiting G dwarfs will not be synchronized, unless they form
with a very large rotational period, which is possible
\citep{MiguelBrunini10}. This difference is due to the different frequency dependencies between the two models. The ratio of Eq.~(\ref{eq:o_cpl}) to Eq.~(\ref{eq:o_ctl}) is $1/(2nQ\tau|n - \Omega|)$, and for an Earth with an initial rotation period of 10 days, the ratio is 9, \ie the CPL model initially de-spins the planet 9 times faster than the CTL model. The dependence on $\Omega$ forces the ratio to increase with time as $\Omega~\rightarrow~n$, and, indeed, for this case the difference in the time to tidally lock is a factor of 50. If Earth formed with no Moon, an initial
rotation period of 3 days, and maintained a constant $Q$ value of 12,
then the CPL model predicts it would become synchronously rotating
within 5~Gyr. The reader is cautioned that extrapolating from the
Earth-Moon frequency to the Earth-Sun frequency could introduce
significant error in the tidal modeling. The CTL's frequency
dependence maintains more similar responses over a wider frequency
range than CPL, and hence the CTL model does not predict such short
tidal locking times for Earth-like planets in the HZ of G dwarfs.

If the assumption of a circular orbit is relaxed, then the evolution
can be significantly more complicated. As demonstrated in
Fig.~\ref{fig:k22b}, planets on higher eccentricity orbits can 
rotate super-synchronously. This aspect
of tidal theory is well-known
\citep{Goldreich66,MurrayDermott99,Barnes08}, but the details depend on the planet. ET models predict that super-synchronous rotation will occur for any $e>0$, but it is far more likely that planets will be locked into spin-orbit resonance such as 1:1, 3:2, or 2:1
\citep{Rodriguez12}. However, the tidal braking process
results in transfer of rotational angular momentum to orbital angular
momentum and can lead to a change in $e$. The sign of the change
depends on the ratio of the rotational and orbital frequencies of both
bodies. Previous results that predict eccentricities always decrease,
such as \cite{Rasio96} and \cite{Jackson08a}, have assumed the planets
had already synchronized.

\begin{figure} 
\begin{center}
\includegraphics[width=0.9\textwidth]{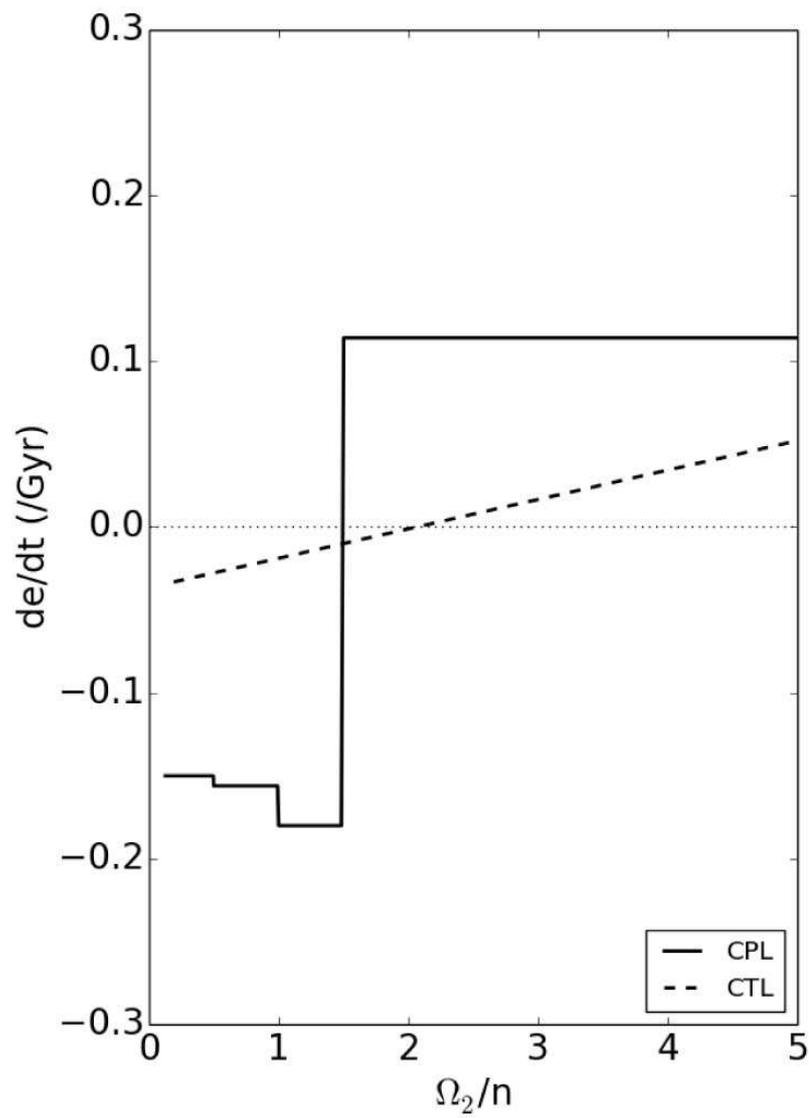} 
\end{center}
\caption{$de/dt$ for an Earth-like planet orbiting a $0.1~\msun$ star
  as a function of the angular frequencies, with $\Omega_2$ being the
  rotational frequency of the planet. The semi-major axis of the orbit
  is 0.05~AU, and the eccentricity is 0.2. Both models predict that
  for fast rotators, $de/dt$ is positive.}
\label{fig:edot}
\end{figure}

The coupled evolution of the rotational frequency and the orbital
eccentricity is a complex function of the dissipation rates and the
distribution of the angular momentum among the spins and orbit. As can
be seen in Eqs.~(\ref{eq:e_cpl}) and (\ref{eq:e_ctl}), the change in
eccentricity can be positive or negative depending on the ratios of
several parameters. In general, $e$ will decay for rotational periods
longer than or similar to the orbital period.  This behavior can be
understood in terms of forces. A body rotating faster than it is
revolving at pericenter will have a bulge that leads the
perturber. This bulge has a gravitational force that accelerates the
perturber in the direction of the motion. This acceleration increases
the tangential velocity at pericenter and hence the eccentricity must
grow. Because the pericenter distance must remain constant, the
semi-major axis must also grow. Thus, the equilibrium rotation period
can grow or shrink depending on the rates of change of $a$ and $e$. As
an example, the value of $de/dt$ as a function of $\Omega/n$ for an
Earth-like planet orbiting $0.1~\msun$~star with $a=0.05$~AU and
$e=0.2$ is shown in Fig.~\ref{fig:edot}. In the CPL model, planets
that rotate faster than $1.5n$ will
experience eccentricity growth, while slower rates lead to
eccentricity decay. In the CTL model the critical frequency ratio
depends on eccentricity. The difference between the two models is due
to the different assumptions of the frequency dependence.

\begin{figure} 
\begin{center}
\includegraphics[width=0.75\textwidth]{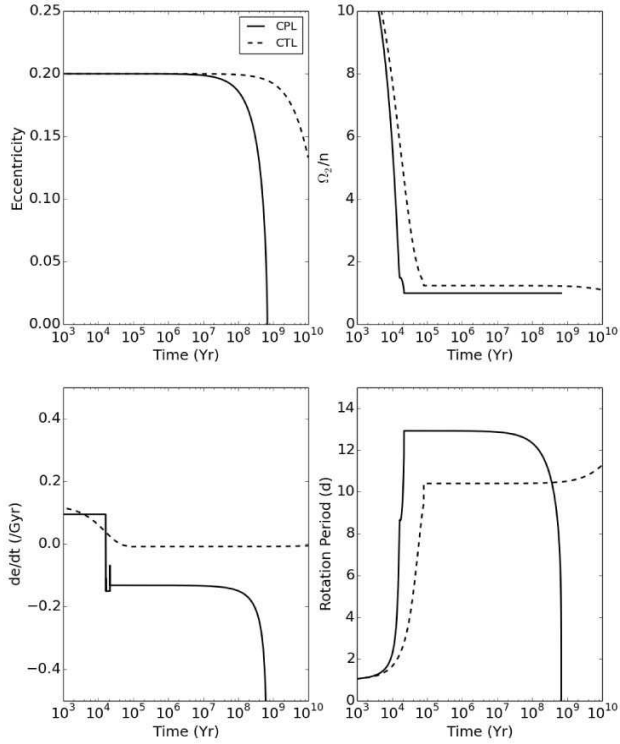} 
\end{center}
\caption{Rotational and orbital evolution of an Earth-like planet
  orbiting a $0.1~\msun$ star at 0.05~AU. The CPL model is represented
  by the solid curve and the CTL by the dashed. Note that after 680
  Myr the CPL model predicts the planet will merge with the host
  star. {\it Top Left:} Eccentricity. {\it Top Right:} The ratio of
  the planet's rotational frequency to mean motion. {\it Bottom Left:}
  The time derivative of the eccentricity. The CPL predicts several
  discontinuities as the rotational frequency evolves. {\it Bottom
    Right:} The planet's rotational period. The CPL model predicts $a$
  will decay rapidly after about 100 Myr, and since the planet is
  rotationally locked its period drops. The CTL model predicts later
  eccentricity damping without much evolution in $a$, and hence the
  rotational period rises.}
\label{fig:eccgrowth}
\end{figure}

Fig.~\ref{fig:eccgrowth} shows the orbital and rotational evolution of
an Earth-like planet orbiting at 0.05~AU. The eccentricity grows
slightly until $n~\approx~1.5\Omega$ at which point it begins to
decay. For many cases, this process can act to delay the
circularization process.  The eccentricity distribution of
radial-velocity-detected exoplanets with orbits that are unaffected by
tidal interaction, \ie those with $a~>~0.2$~AU, has a
mean\footnote{See http://exoplanets.org} of $\sim~0.3$. While the vast
majority of these exoplanets is expected to be gaseous, we should
nonetheless expect many terrestrial exoplanets to possess large
eccentricities and those that are tidally locked will rotate
super-synchronously. However $e$ will likely also tidally evolve, and
hence the equilibrium period will change with time, too.  As
rotational evolution tends to be more rapid than orbital evolution,
tidal locking generally occurs prior to orbital circularization. If
$de/dt > 0$, then the planet can be tidally locked and evolving to
ever shorter rotation periods. In most cases, $e$ decays with time and
hence the rotational period slows, possibly in a series of sudden
transitions between spin-orbit resonances, a fascinating scenario for
the biospheres of inhabited exoplanets.

\begin{figure} 
\begin{tabular}{cc}
\includegraphics[width=0.49\textwidth]{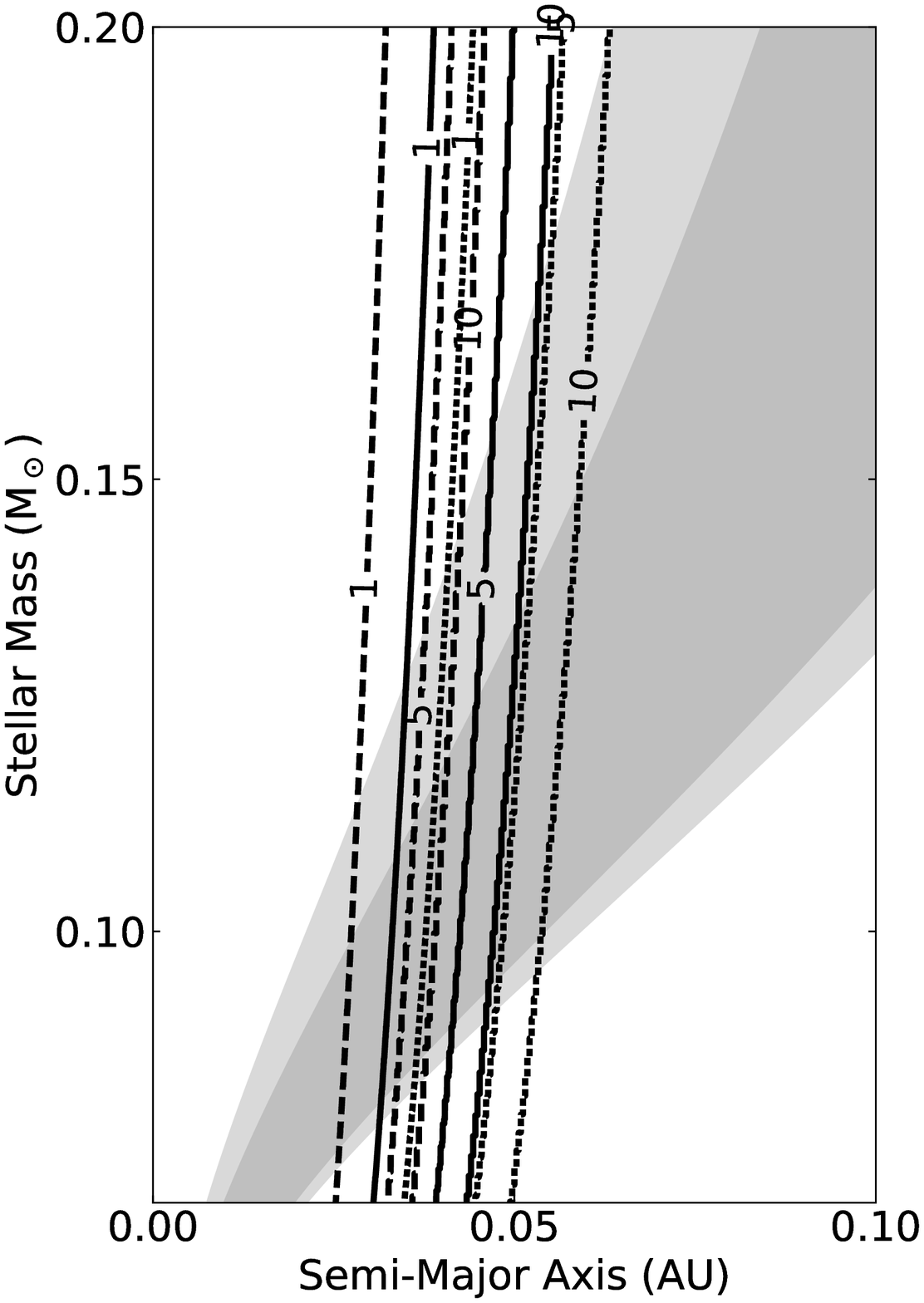} &
\includegraphics[width=0.49\textwidth]{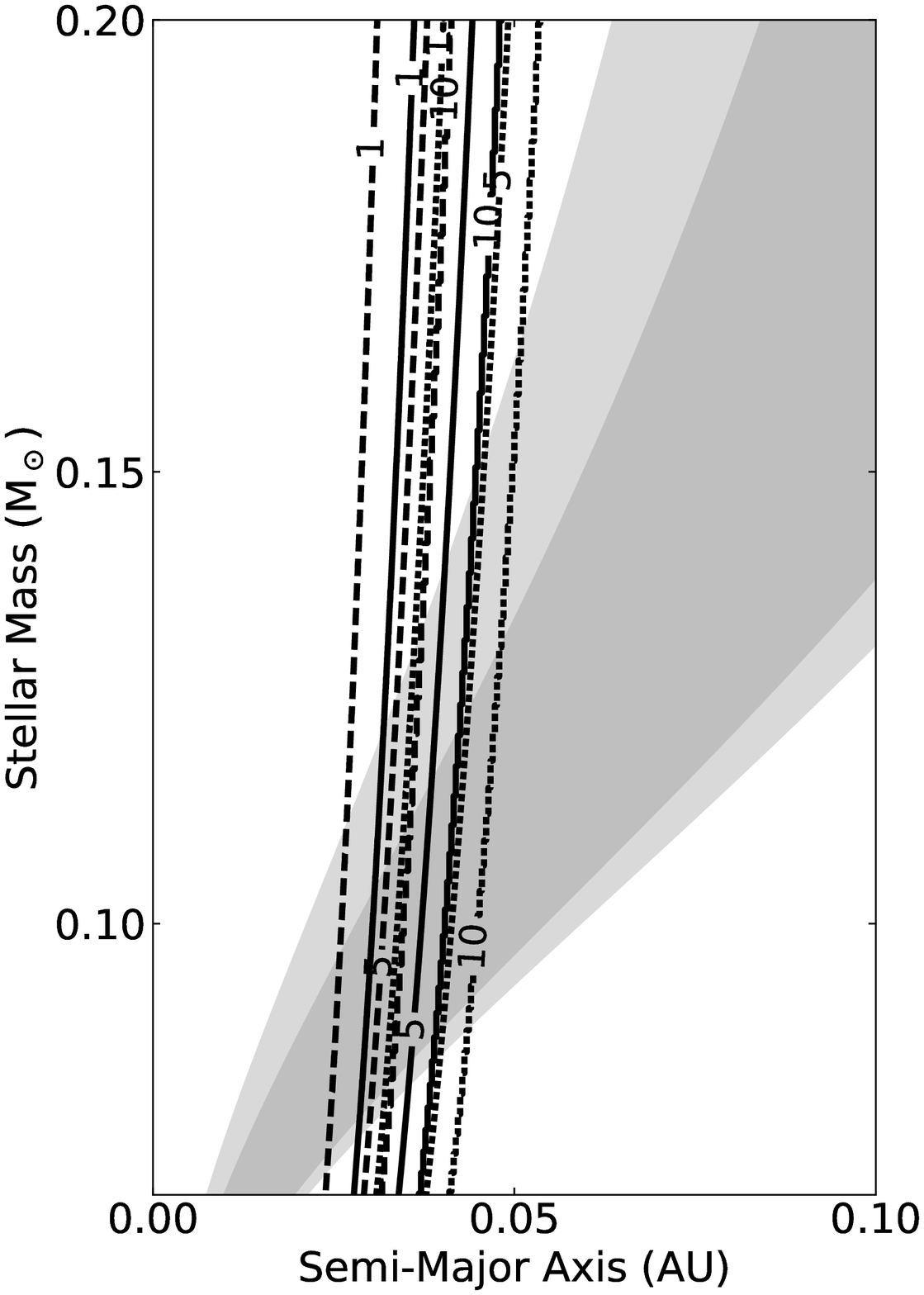}\\
\end{tabular}
\caption{The time in Gyr for an Earth-like planet with an
  initial eccentricity of 0.3 to circularize. The format is the same
  as Fig.~\ref{fig:circtime}. The left panel is the CPL model, right
  is CTL.}
\label{fig:circtime}
\end{figure}

Figure~\ref{fig:circtime} shows the time required for the eccentricity
of an Earth twin planet to drop from 0.3 to 0.01, the circularization
timescale $T_{circ}$, for a range of stellar masses. The initial
period is 1 day, and the initial obliquity is $23.5^\circ$ in these
calculations. The contour lines denote $T_{circ}$ in Gyr. For both
models, Earth-like planets orbiting stars less massive than
$0.1~\msun$ will likely be on circular orbits and in synchronous
rotation. Planets with initially fast rotation rates, or strong
perturbations from other planets might not be synchronous rotators
around stars of any mass.

\subsection{{\it TESS} Projections}

The results of the previous section provide a foundation to interpret
future discoveries from transit \citep{BertaThompson15,Gillon16},
radial velocity \citep{AngladaEscude16}, and astrometric data
\citep{Malbet16}. In the immediate future, the {\it Transiting
  Exoplanet Survey Satellite} ({\it TESS}) will likely discover dozens
to hundreds of potentially habitable worlds, a handful of which could
be amenable to atmospheric characterization by {\it JWST}
\citep{Sullivan15,Barnes15_hite}. In this section, I calculate the
tidal evolution of the predicted sample provided by \citep{Sullivan15},
which convolved realistic models for the detector, stellar models and
variability, planet occurrence rates from {\it Kepler}, etc., to
produce a hypothetical catalog of detected exoplanets and their
properties. I find that nearly every potentially habitable planet it
will discover will be tidally locked.

\begin{figure} 
\includegraphics[width=\textwidth]{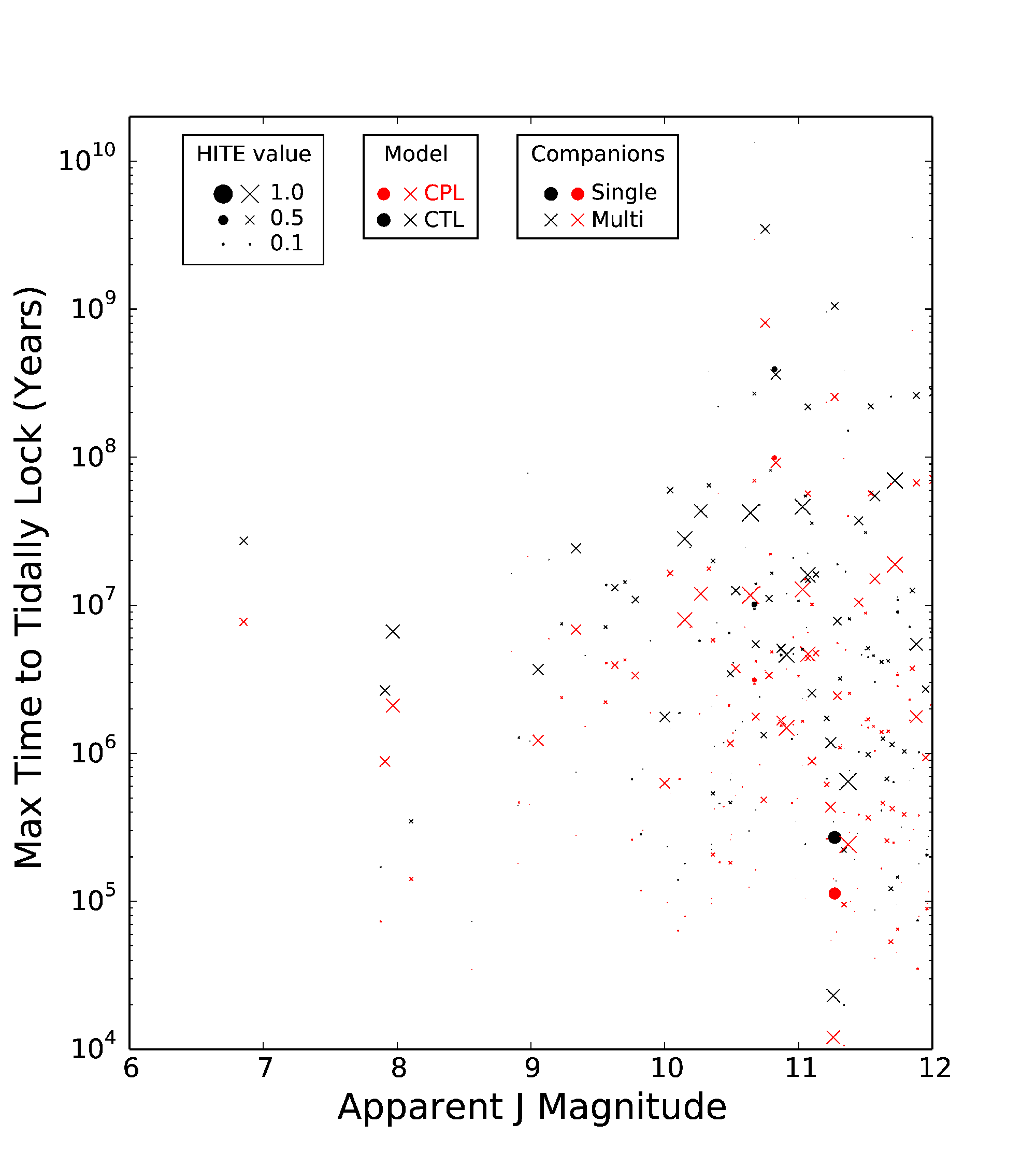}
\caption{Maximum timescale for tidal locking, HITE values and apparent J band magnitudes of potentially habitable planets predicted for {\it TESS} from \cite{Sullivan15}. The tidal locking time assumes a tidal $Q$ of 100, and initial rotation periods and obliquities of 8 hr and $60^\circ$, respectively. The circles correspond to planets for which only one planet is detected, x's correspond to planets in multiplanet systems. Dot and x sizes are proportional to the HITE value.}
\label{fig:tess}
\end{figure}

Fig.~\ref{fig:tess} shows the results of these simulations for the
``long'' cases, i.e. these represent a maximum timescale for these
planets to tidally lock, assuming they are terrestrial. Nearly every
planet tidally locks within 1 Gyr for both models. Table 2 lists the
planets that are predicted to be single (at least in terms of those
that can be discovered), and Table 3 (available in the Online Supplement) lists those planets that would be
known to be in multiple planet systems. In principle, any of these
predicted planets could be in a high eccentricity state, but those
that are known to be in multiple systems may be more likely to have
eccentricities that are sufficiently perturbed by companions to force
the rotation rate into a spin-orbit resonance.

\section{Discussion and Conclusions\label{sec:discussion}}

\begin{figure} 
\includegraphics[width=\textwidth]{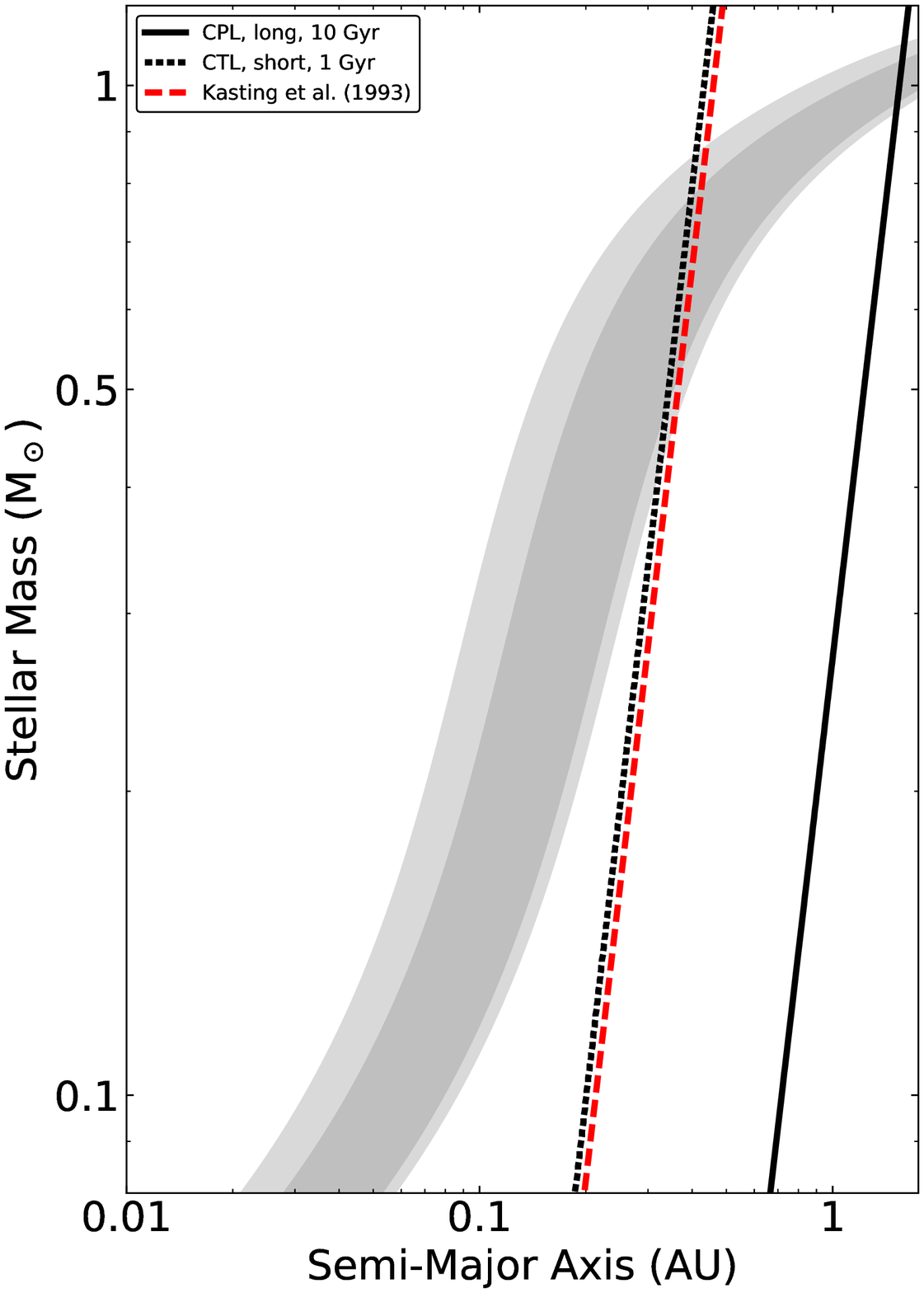}
\caption{A comparison of the HZ and the extreme limits for tidal locking obtained in this study and \cite{Kasting93}. The grey regions are the HZ as shown in previous figures. The dashed red curve is the tidal lock radius for the model used by \cite{Kasting93} (Earth-mass, 13.5 day period, 4.5 Gyr age). The dotted black line is for a 1 Earth-mass planet with an initial spin period of 8 hr and obliquity of $60^\circ$ and 1 Gyr. The solid black line is for the CPL.}
\label{fig:hzplot}
\end{figure}

Previous studies of the tidal evolution of habitable exoplanets
identified those orbiting M dwarfs as the most likely, or perhaps
only, candidates for synchronous rotation, but this study has shown
that the CPL model predicts that planets orbiting in the HZ of
solar-mass stars may also be synchronously
rotating. Fig.~\ref{fig:hzplot} compares the results of
\cite{Kasting93} to the extreme contours of Fig.~\ref{fig:survey} and
shows that the \cite{Kasting93} prediction is close to the innermost
``tidal lock radius'' found in this study. As recent simulations of
planet formation have found that initial rotation periods between 10
hours and 40 days are approximately equally likely
\citep{MiguelBrunini10}, the tidal lock radius of \cite{Kasting93} is
now seen to be based on a very short initial rotation period. The
above calculations only considered initial rotation periods up to 10
days, and so those results are still conservative -- tidal locking may
be even more likely than presented here, let alone than in
\cite{Kasting93}.

As astronomers develop technologies to directly
image potentially habitable planets orbiting FGK dwarfs
\citep[\eg][]{Dalcanton15}, they must be prepared for the possibility that
planets orbiting any of them may be tidally locked. Such a rotation
state can change planetary climate, and by extension the reflected
spectra. 3D models of synchronously rotating habitable planets should
be applied to planets orbiting K and G dwarfs in addition to Ms. While
not explicitly considered here, habitable worlds orbiting brown dwarfs
and white dwarfs are even more likely to be synchronous rotators, but
their potential habitability is further complicated by the luminosity
evolution of the central body \citep{BarnesHeller13}.

Current technology favors the detection of habitable planets that
orbit close to their host star, and hence the planets are more likely
to be tidally locked than Earth. This expectation is borne out by the
simulations presented in $\S$~\ref{sec:results}. Proxima b will be a
prime target for future observations, and it is almost assuredly
tidally locked, and it is highly likely that all potentially habitable
\tess~planets will be tidally locked. The {\it JWST} telescope may be
able to spectroscopically characterize the atmosphere of a few
\tess~planets, and so any biosignature search should consider the role
of tidal locking.

Both the CPL and CTL models predict that fast rotation can cause
eccentricity growth, but in most cases such growth is modest, acting
primarily to delay circularization. This delay can be significant and
thus studies that use current orbits to estimate tidal parameters
\citep{Jackson08a,Matsumura10} may incorrectly constrain the
dissipation rates because they did not take into account the role of
rotation. The initial rotation rate is unknown in the vast majority of
cases, and so constraining the orbital history of a planet that is
currently on a circular orbit and synchronously rotating is very
challenging.

For $e~\lsim~0.12$, the exoplanet is likely to
be trapped in a 1:1 spin-orbit resonance \citep{Rodriguez12}, as is the case for
Earth's Moon with $e=0.05$. However, the position of the host body
will not be fixed in the sky, but will librate due to the changing
orbital angular velocity and the torques that try to drive the
tidally-locked body into a super-synchronous state \citep{Makarov16}.
Given the propensity of M dwarf planets to be multiple
\citep{BallardJohnson16}, planets in their HZs will likely have nonzero
eccentricities due to mutual gravitational perturbations. Thus, the
climates of librating planets should be modeled to determine how
libration affects the limits of habitability.

The role of rotation is critical for planetary habitability, and could
also be used to discriminate between tidal models. The results
presented here show that a more thorough examination of the
differences between the CPL and CTL model admits
possibilities that could significantly impact our search for life in
the universe. As astronomers have focused on discovering habitable
exoplanets, a natural synergy will emerge between measuring rotation
rates and constraining tidal evolution. When numerous planetary
rotation rates of habitable exoplanets have been measured, the nature
of tidal evolution may also be revealed.

\begin{acknowledgements}
This work was supported by NSF grant AST-1108882 and by the NASA
Astrobiology Institute's Virtual Planetary Laboratory under
Cooperative Agreement number NNA13AA93A. I thank Brian Jackson, Ren{\'
e} Heller, J{\' e}r{\' e}my Leconte and two anonymous referee for
their comments that greatly improved the content and clarity of this
manuscript. This research has made use of the NASA Exoplanet Archive,
which is operated by the California Institute of Technology, under
contract with the National Aeronautics and Space Administration under
the Exoplanet Exploration Program. I also thank GitHub for free
hosting of the source codes used in this manuscript, available at
https://github.com/RoryBarnes/TideLock.
\end{acknowledgements}

\bibliography{bib}

\newpage
\begin{center}Table 2: Orbital and Physical Parameters for Projected Isolated Exoplanets from {\it TESS} \\ 
\begin{tabular}{cccccccccccccc}
  \hline
  ID & $R_*$ & $M_*$ & $P_{orb}$ & $R_p$ & $M_p$ & $a$ & HITE & $T_{CPL}^{short}$ & $T_{CPL}^{\oplus}$ & $T_{CPL}^{long}$ & $T_{CTL}^{short}$ & $T_{CTL}^{\oplus}$ & $T_{CTL}^{long}$ \\
   & (R$_\odot$) & (M$_\odot$) & (d) & (R$_\oplus$) & (M$_\oplus$) & (AU) & & (Myr) & (Myr) & (Myr) & (Myr) & (Myr) & (Myr)\\
  \hline
          105 &   0.14 &   0.14 &   7.58 &   1.68 &   6.70 &   0.04 & 0.66 &  0.003 &  0.010 &  0.113 &  0.003 &  0.084 &  0.270\\
          606 &   0.56 &   0.58 &  42.19 &   1.82 &   8.99 &   0.20 & 0.26 & 11.713 & 11.629 & 99.198 & 11.713 & 136.965 & 393.163\\
          724 &   0.39 &   0.40 &  15.50 &   2.42 &  25.29 &   0.09 & 0.02 &  0.126 &  0.245 &  2.284 &  0.126 &  2.307 &  7.062\\
          855 &   0.35 &   0.35 &  17.56 &   1.87 &   9.86 &   0.09 & 0.27 &  0.210 &  0.344 &  3.130 &  0.210 &  3.327 & 10.114\\
          946 &   0.29 &   0.29 &  12.55 &   2.09 &  14.79 &   0.07 & 0.15 &  0.030 &  0.094 &  0.913 &  0.030 &  0.849 &  2.634\\
         1344 &   0.33 &   0.33 &  11.52 &   2.43 &  25.52 &   0.07 & 0.01 &  0.012 &  0.073 &  0.722 &  0.012 &  0.647 &  2.019\\
         1450 &   0.28 &   0.27 &  16.50 &   2.33 &  22.02 &   0.08 & 0.12 &  0.178 &  0.309 &  2.841 &  0.178 &  2.947 &  8.984\\
         1515 &   0.26 &   0.25 &   7.18 &   2.04 &  13.57 &   0.05 & 0.03 &  0.003 &  0.009 &  0.104 &  0.003 &  0.075 &  0.243\\
         1753 &   0.17 &   0.15 &   9.36 &   2.11 &  15.24 &   0.05 & 0.27 &  0.000 &  0.028 &  0.295 &  0.000 &  0.242 &  0.764\\
         1899 &   0.43 &   0.45 &  15.59 &   1.70 &   6.86 &   0.09 & 0.08 &  0.105 &  0.199 &  1.853 &  0.105 &  1.878 &  5.742\\
\end{tabular}
\end{center}

\end{document}